\documentclass[twocolumn,secnumarabic,amssymb, nobibnotes, aps, prx,superscriptaddress]{revtex4-2}

\usepackage[paperwidth=210mm,paperheight=297mm,centering,hmargin=2cm,vmargin=2.5cm]{geometry}
\usepackage{amsmath} % Advanced math typesetting
\usepackage[english]{babel} % Change hyphenation rules
\usepackage{graphicx} % Add pictures to your document
\usepackage{listings} % Source code formatting and highlighting
\usepackage{soul}
\usepackage{color}
\usepackage{lipsum}
\usepackage{float}

\begin{document}

\title{Electrostatics and viscosity are strongly linked\\in concentrated antibody solutions}

\author{Fabrizio Camerin}
\email[Corresponding author:]{ fabrizio.camerin@fkem1.lu.se}
\affiliation{Division of Physical Chemistry, Department of Chemistry, Lund University, Lund, Sweden}

\author{Marco Polimeni}
\affiliation{Division of Physical Chemistry, Department of Chemistry, Lund University, Lund, Sweden}

\author{Anna Stradner}
\affiliation{Division of Physical Chemistry, Department of Chemistry, Lund University, Lund, Sweden}
\affiliation{LINXS Institute of Advanced Neutron and X-ray Science, Lund University, Lund, Sweden}

\author{Emanuela Zaccarelli}
\email[Corresponding author:]{ emanuela.zaccarelli@cnr.it}
\affiliation{CNR Institute of Complex Systems, Uos Sapienza, Piazzale Aldo Moro 2, 00185 Roma, Italy}
\affiliation{Department of Physics, Sapienza University of Rome, Piazzale Aldo Moro 2, 00185 Roma, Italy}

\author{Peter Schurtenberger}
\email[Corresponding author:]{ peter.schurtenberger@fkem1.lu.se}
\affiliation{Division of Physical Chemistry, Department of Chemistry, Lund University, Lund, Sweden}
\affiliation{LINXS Institute of Advanced Neutron and X-ray Science, Lund University, Lund, Sweden}

\date{\today}

\begin{abstract}
\noindent

Monoclonal antibodies are among the most promising therapeutic agents in modern medicine, yet their formulation into high-concentration solutions for subcutaneous self-administration poses a major challenge. A key obstacle is the marked increase in viscosity often observed under these conditions.
To gain deeper insights into this phenomenon, coarse-grained models derived from soft matter physics have been widely employed. However, these models have yet to be fully leveraged for analyzing the rheological collective properties of such systems.
In this study, using molecular dynamics simulations, we directly compute the antibody solution viscosity by starting from commonly used models in which electrostatic interactions are treated through effective screened Coulomb potentials. We demonstrate that this approach fails to reproduce experimental evidence and we show, by analyzing stress correlations in the system, that it is necessary to treat the heterogeneously charged domains explicitly, also including counterions and salt ions, and to properly account for the long-ranged nature of Coulomb interactions.  By thoroughly analyzing the microscopic structure of the system, we further reveal the presence of transient strongly correlated antibodies which would not be present if charges were treated implicitly, thus pointing to a prominent role of electrostatics in determining the increase in viscosity at high concentrations.
By taking advantage of our realistic treatment, new approaches can be devised to ensure that antibody solutions exhibit the desired characteristics for their intended broad use and effective deployment.
\end{abstract}

\maketitle
%\tableofcontents

\section*{Introduction}

Many of the outstanding breakthroughs in pharmacology and medicine require concerted actions by the scientific community to overcome shortcomings that would otherwise prevent their widespread use. Challenges may relate to different aspects of the newly developed products and can range from the stability of the formulation over time~\cite{le2020physicochemical,nejadnik2018postproduction}, to issues in the large-scale production~\cite{sarkis2021emerging}, or to the method of administration to patients~\cite{bittner2018subcutaneous,mitragotri2014overcoming}.

A tangible example is offered by monoclonal antibodies that recent medical research places at the forefront in modern cancer treatments, for their versatility in targeting virtually any substance by properly adapting their antigen binding site~\cite{scott2012monoclonal}. Besides, they are already used as an efficient medication for many immunological and allergic diseases~\cite{singh2018monoclonal,castelli2019pharmacology}. %Their effectiveness also appears to be granted, since bla bla.
A major issue for exploiting the potentialities of these molecules, rather than concerning their development and production, is physical in origin. In fact, one of the methods that has been increasingly gaining ground for the administration of antibodies is by using simple syringes for subcutaneous self-injection~\cite{viola2018subcutaneous,davis2024subcutaneous}. This method could be leveraged autonomously by patients and would be as effective as intravenous infusion, which would instead require hospitalization and entails generically higher costs. However, the very limited volume that can be injected and the large number of antigen binding sites usually needed demand solutions to be highly concentrated. 
The enhancement of viscosity and/or turbidity of the solution that is typically observed in these conditions severely limits the applicability of antibodies as self-injectable, safe, and efficient \textit{biologics}~\cite{filipe2012immunogenicity,roberts2014protein,lundahl2021aggregation,pham2020protein}.

This problem can only be addressed through an interdisciplinary approach, in which the biological and structural characteristics of antibodies are considered alongside the physical understanding of the increased viscosity.
Throughout the years, this undesirable effect has often been attributed to the presence of clusters that form in solution as a result of equilibrium interactions between antibodies~\cite{lilyestrom2013monoclonal,skar2023using}.
Regardless of the actual primary reason that leads to the increase in viscosity, it is of utmost importance to develop a comprehensive toolbox that can effectively address the underlying mechanisms of this phenomenon. In this context, the use of descriptors that have been typically employed for globular proteins, such as the second virial coefficient or the diffusion interaction parameter~\cite{connolly2012weak,dear2017contrasting,saito2012behavior,godfrin2016effect}, should be appropriately justified and substantiated. 
%In fact, they are only defined in the dilute regime, posing questions on their reliability when used for predictions at high concentrations.
Most importantly, a treatment based on such spherically symmetric parameters would completely ignore the characteristic Y-shape of antibodies and the effect of specific directional interactions. As a matter of fact, to provide a microscopic interpretation of macroscopic experimental observables, an appropriate coarse-graining procedure is required, avoiding gross simplifications with idealized models and too refined atomistic treatments, for which an overall understanding of the collective behavior would be not achievable~\cite{stradner2020potential}. The use of appropriate modeling and simulations would allow to predict in advance the behavior of antibodies, to deepen the understanding of experimental outcomes and to ultimately guide the design of new antibodies and antibody formulations with desired properties. 

A promising avenue relies on mapping the physically relevant features of the biological object into monomer-based models inspired by anisotropic colloids and polymers~\cite{stradner2020potential}. In this respect, several works in the field proposed a large number of models aimed at reproducing specific experimental properties, from the compressibility of the system to the excluded volume of the antibody, or the static structure factors~\cite{calero2016coarse,skar2023using,dear2019x,polimeni2024multi}. % or the diffusion of the antibodies.
Nonetheless, a univocal consensus has not been reached yet, and different models are being used by different research groups. For example, from a structural perspective, the size that the colloidal molecule should retain to best reproduce the excluded volume of the antibodies is still debated, with bead-based models employing different numbers of beads being equally employed
%and 6, 9, 12 or more bead-based models are all equally employed
~\cite{skar2019colloid, calero2016coarse,blanco2019evaluating,polimeni2024multi}. 

\begin{figure*}
\centering
\includegraphics[width=\textwidth]{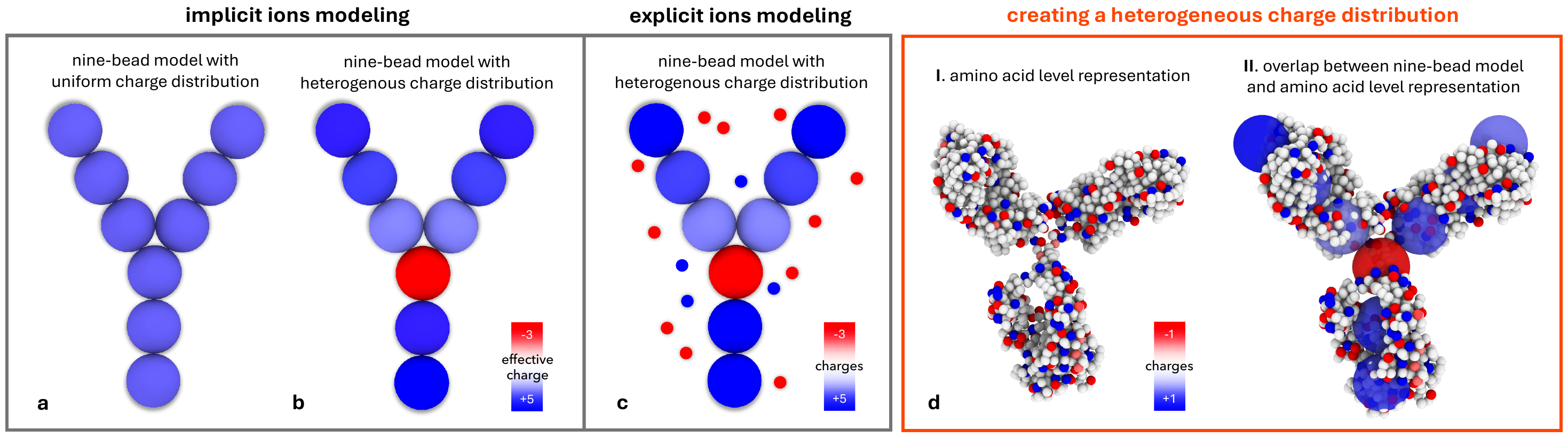}
\caption{\small \textbf{Antibody coarse-grained modeling.} Implicit ions antibody models in which every bead interacts according to a Yukawa effective potential, according to (a) a uniform and (b) a heterogeneous charge distribution, respectively.
(c) Explicit ions antibody model with heterogeneous charge distribution and representation of (not-in-scale) ions and counterions. (d) Mapping of the charges starting from (I) the amino acid representation, (II) overlapping it with the coarse-grained 9-bead model, for then leading to the final coarse-grained model adopted (either b or c).
The color coding reflects the charge assigned to each bead, as from the color bar. Dark shading around the beads of the colloidal model represents the short-range van der Waals attraction.}
\label{fig:models}
\end{figure*}

% An aspect that has been often overlooked in the modeling is the role of heterogeneously charged domains, of which proteins and antibodies are particularly abundant. As an example, electrostatics is expected to play a crucial role for describing correctly the proximity to a liquid-liquid phase separation or to justify the formation of strongly interacting transient clusters~\cite{godfrin2016effect,notarmuzi2024features,liu2005reversible,skar2019colloid,stradner2004equilibrium}. Since both these phenomena have been related to the increase in viscosity in colloidal systems, charges represent an essential ingredient for investigating concentrated antibody solutions. 
% However, with some recent relevant exceptions~\cite{notarmuzi2024features}, most of previous work has only led to qualitative indications about the role of charges and has been based on the use of (isotropic) effective potentials.
An aspect that has been often overlooked in the modeling is the role of heterogeneously charged domains, of which proteins and antibodies are particularly abundant. These are expected to play a crucial role for describing correctly the proximity to a liquid-liquid phase separation or to justify the formation of strongly interacting transient clusters~\cite{godfrin2016effect,liu2005reversible,skar2019colloid,stradner2004equilibrium}. Since both these phenomena have been related to the increase in viscosity in colloidal systems, charges represent an essential ingredient for investigating concentrated antibody solutions. 
However, most of previous work has only led to qualitative indications about the role of electrostatics and has been based on the use of (isotropic) effective potentials.
For instance, there have been attempts to use patchy colloidal models to include interactions between oppositely charged domains and for investigating their effect on the antibody self-assembly, especially related to the formation of branched polymer-like structures~\cite{skar2019colloid}.
In other cases, charges have been treated by assigning an effective Yukawa potential to each bead, incorporating the screening of the Coulomb potential by counterions and salt ions, and thus determining the overall charge as a fit parameter to compare to static structure factors~\cite{dear2019x}. 
%Few models have also considered beads with specific localized attractions with charges assigned in an implicit manner~\cite{dear2019x}.
Few models have also considered beads with specific localized attractions~\cite{dear2019x}, or with additional domain beads aimed at reproducing a close-to-atomic level of detail for electrostatic interactions~\cite{shahfar2021toward,forder2023simulation}, with charges always treated by means of implicit screened Coulomb potentials.
%However, charges have always been treated by means of implicit screened Coulomb potentials.
All these models have been tested for reproducing static features of the system and, in most cases, a reasonable qualitative agreement has been achieved. 

Surprisingly, however, only a very few models have been used for directly calculating dynamical properties with molecular dynamics simulations~\cite{lai2021calculation,wang2018structure}, despite the prediction of the viscosity in concentrated antibody solutions being one of the main research questions in the field. Furthermore, direct comparison with experimental measurements reveals the limitations of the models employed.
%Surprisingly, however, none of the models has been used for directly calculating dynamical properties with molecular dynamics simulations, despite the prediction of the viscosity in concentrated antibody solutions being one of the main research question in the field. 
In the remaining cases, the dynamical and rheological properties of antibodies in crowded environments have been studied by means of experiments alone or, theoretically, by applying phenomenological constitutive rheological equations, such as the so-called Mooney relation, or by means of empirical relations~\cite{chowdhury2020coarse,dear2019x,gulotta2024combining, skar2019colloid,kastelic2018a, Chowdhury2023c, Hung2019, Tomar2016, Woldeyes2019, Yadav2012, Brudar2024, Arora2016, Salinas2010}.
%Therefore, in light of these considerations, there is the need to identify the specific limitations of the current methodology and to propose a more refined and comprehensive protocol for advancing the state-of–the-art.
Therefore, in light of these considerations, there is the need to identify specific shortcomings in the current methodology and to develop a more refined and comprehensive protocol to advance the state-of-the-art.

\begin{figure}[b]
\centering
\includegraphics[width=0.48\textwidth]{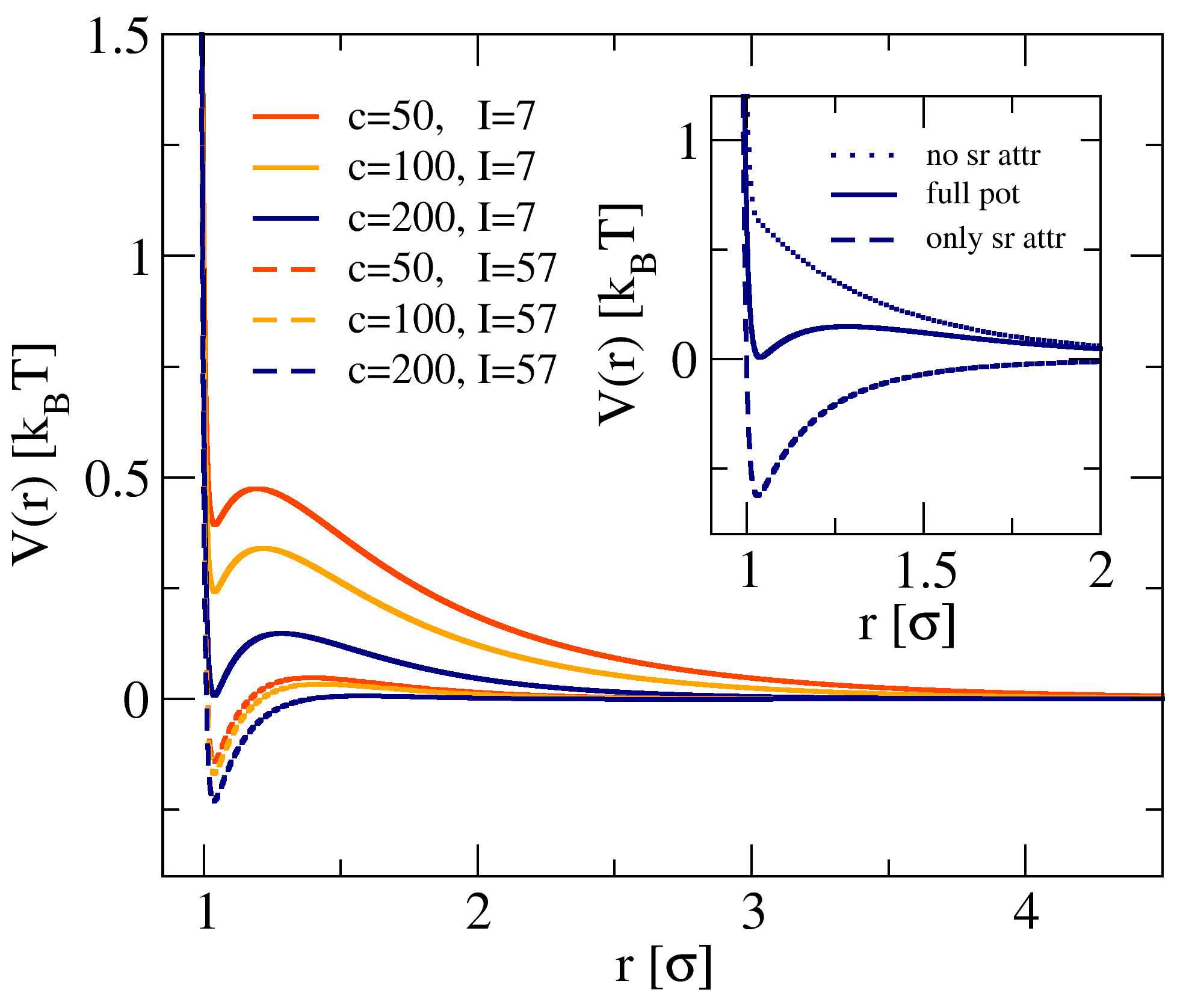}
\caption{\small \textbf{Effective interaction potential.} Implicit ions model bead-bead effective interaction potential V(r) as a function of the distance between bead-bead distance $r$ in units of the bead diamter $\sigma$ for ionic strengths $I=7$ mM (full lines) and $I=57$ mM (dashed lines) for different antibody concentrations $c$, as from the legend. Inset: Overall potential (full pot), potential with no short-range attraction included (no sr attr), and potential with only a short-range attraction included (only sr attr) for the implicit ions model with a uniform charge distribution, for $c=200$ mg/ml and $I=7$ mM.
}
\label{fig:potentials_yuk}
\end{figure}

\begin{figure*}[t!]
\centering
\includegraphics[width=\textwidth]{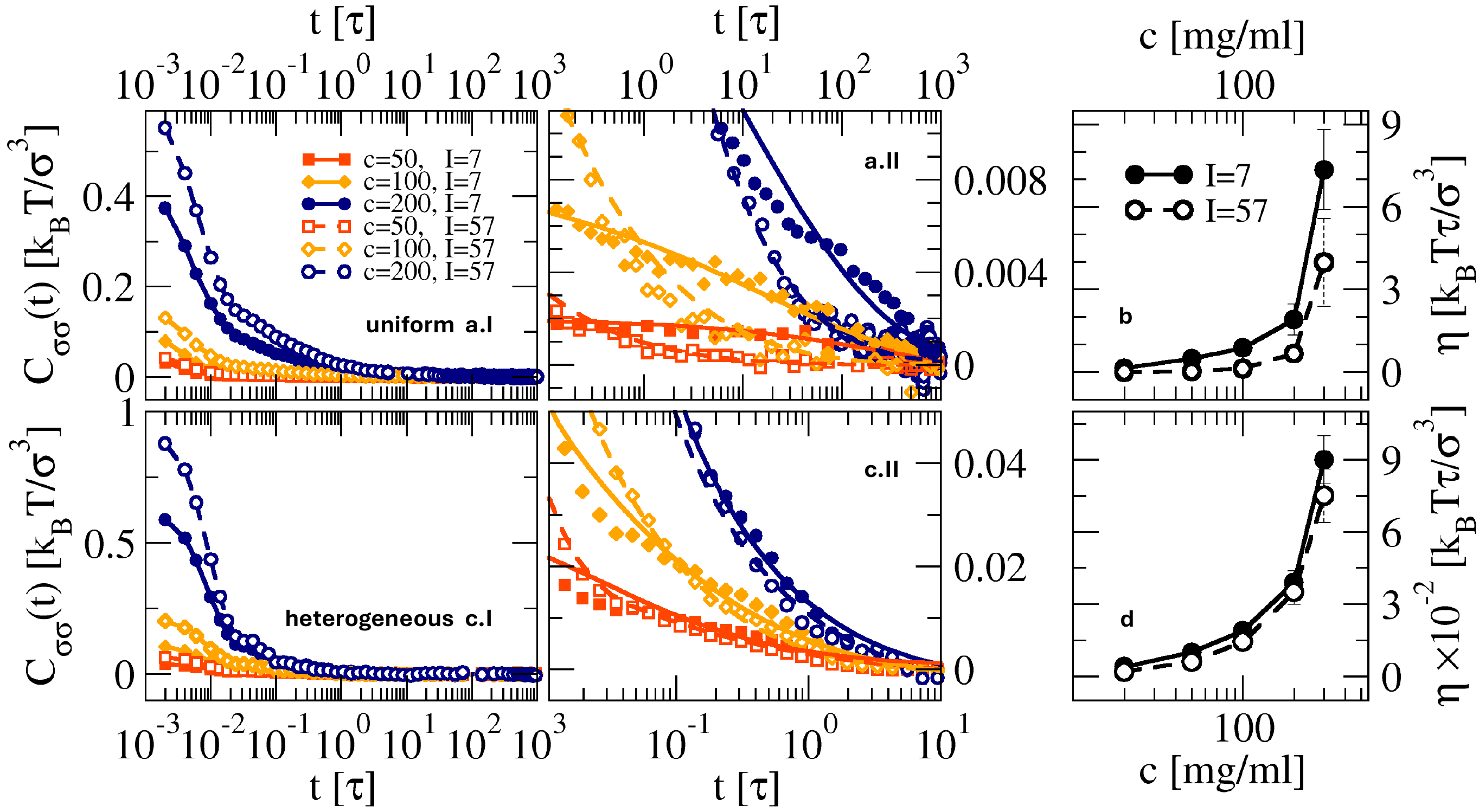}
\caption{\small \textbf{Rheological response for the implicit ions models.} (a,c) Stress autocorrelation function $C_{\sigma\sigma}(t)$ as a function of simulation time $t$ for implicit ions models with uniform and heterogeneous charge distributions, respectively, for ionic strengths $I=7$ mM (filled symbols) and $I=57$ mM (empty symbols) for different antibody concentrations $c$ (expressed in mg/ml), as from the legend. Panels II show an enlargement of $C_{\sigma\sigma}(t)$ for the long-time correlations. Full and dashed lines are guides-to-the-eye for $I=7$ mM and $I=57$ mM, respectively. (b,d) Viscosity $\eta$ as a function of the antibody concentration $c$ (expressed in mg/ml) for implicit ions models with uniform and heterogeneous charge distributions, respectively, for ionic strengths $I=7$ mM (full line) and $I=57$ mM (dashed line) and relative error bars.%\fc{check norm for nonunif}
}
\label{fig:stress_visc_yuk}
\end{figure*}

\begin{figure*}[t!]
\centering
\includegraphics[width=\textwidth]{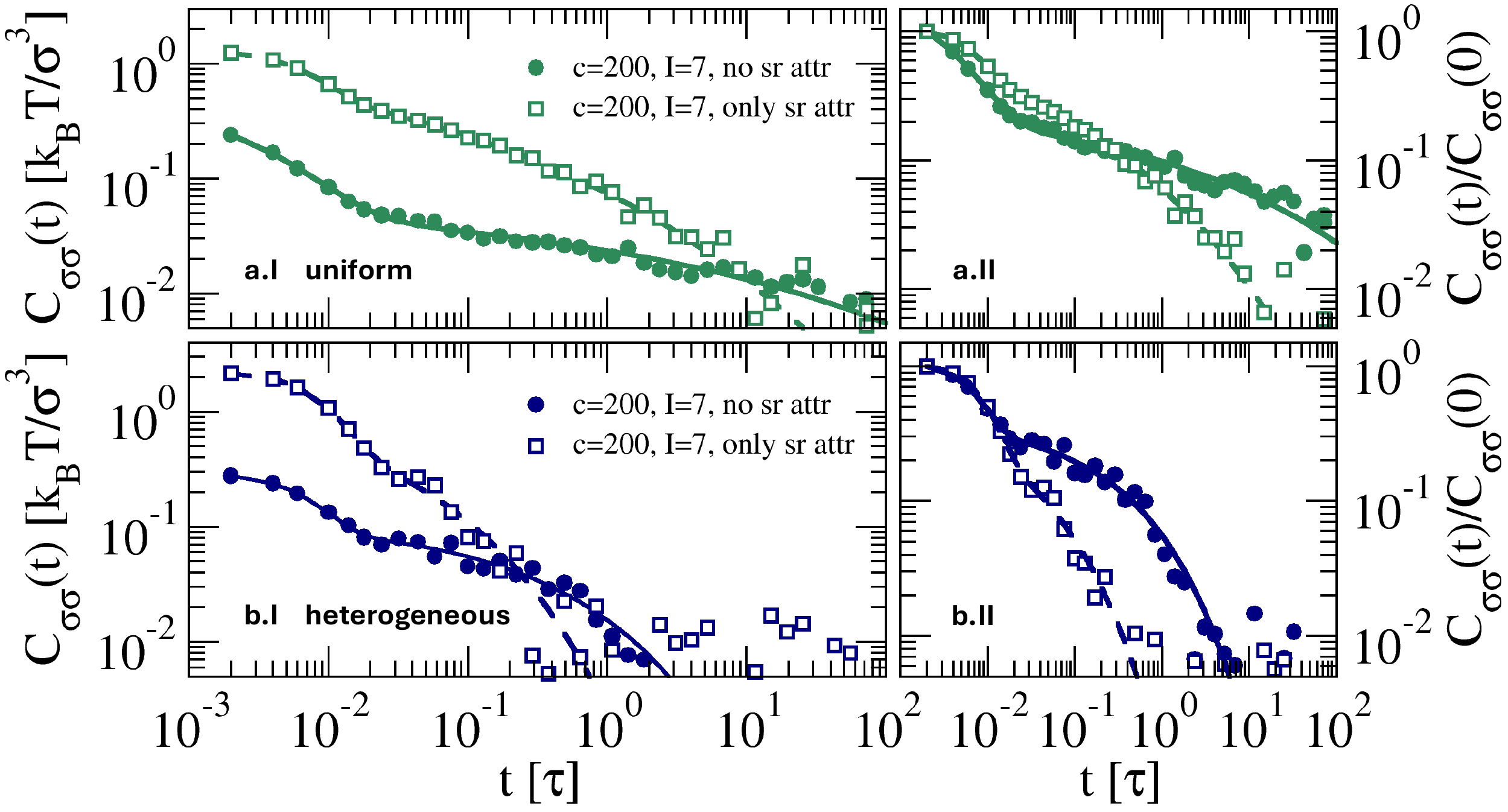}
\caption{\small \textbf{Different contributions to $C_{\sigma\sigma}(t)$ for implicit ions models.} Stress autocorrelation function $C_{\sigma\sigma}(t)$ as a function of simulation time $t$ for implicit ions models with (a) uniform and (b) heterogeneous charge distribution, respectively, in case no short-range attraction (full circles, no sr attr) or only short-range attraction (empty squares, only sr attr) terms are included in the overall bead-bead interaction potential for $c=200$ mg/ml and ionic strength $I=7$ mM. Panels II show the stress autocorrelation function $C_{\sigma\sigma}(t)$ normalized by its initial value $C_{\sigma\sigma}(0)$ at $t=0$. Full and dashed lines are guides-to-the-eye.
}
\label{fig:attr_noattr_yuk}
\end{figure*}

In this work, we establish the fundamental role of charges in shaping the dynamical collective behavior of antibody solutions. By investigating an antibody with an overall repulsive charge at distinct ionic strength conditions, we determine that anisotropic colloidal models are necessary but not sufficient for describing correctly the static \textit{and} the dynamic response of such systems. In particular, through the direct calculation of the zero-shear viscosity, we find that, when charged domains are treated in an implicit fashion by using effective screened Coulomb interactions, the resulting viscosity does not qualitatively agree with experiments. 
%By dissecting the overall interaction potential, we gain insights on the way the system decorrelates its stress, understanding which contributions are not accounted correctly for the study of collective rheological properties.
By dissecting the overall interaction potential, we gain insights into how the system decorrelates stress, enabling us to identify contributions that are not adequately included for correctly describing the collective rheological properties of the system.
Therefore, by combining simulations of antibodies at different scales, we propose a new model where charge heterogeneity is taken into account and, crucially, counterions and salt ions are explicitly incorporated.  We demonstrate for a prototype antibody that these constitute the key discriminating factors for correctly reproducing the behavior of the viscosity observed experimentally. It is only by exploiting an adequate model which incorporates all essential physical ingredients, that a substantial advance in the comprehension of concentrated antibody solutions can be achieved.

%\clearpage
%\newpage
\section*{Results and discussion}

%\subsection{Limited validity of implicit charges models}
\subsection*{Limited validity of models with implicit ions: uniform and non-uniform charge distributions}

We investigate the collective behavior of an antibody IgG1 by making use of anisotropic coarse-grained models, see Figure~\ref{fig:models}. The antibody presents an overall positive charge as evinced from the electrostatic iso-potential surfaces shown in Figure S2.
%, whose representation is given in Figure~\ref{fig:models}(a). 
Its solution properties have been systematically characterized experimentally by some of us~\cite{gulotta2024combining}. 
We start by using the simplest possible description, that was recently shown to be able to correctly reproduce the solution structure factor as a function of antibody and salt concentration~\cite{polimeni2024multi}. In this approach, charges are treated implicitly, so that beads interact according to an effective screened Coulomb (Yukawa) potential, and they are uniformly spread on the antibody beads, as from the representation given in Figure~\ref{fig:models}(a). The electrostatic interaction is additionally complemented by a short-range repulsive potential to account for the beads' excluded volume in a quasi-hard sphere fashion, and by a weakly attractive potential for mimicking short-range van der Waals and hydrophobic interactions. More information on the model is given in the Materials and Methods section. The overall bead-bead interaction potential is reported in Figure~\ref{fig:potentials_yuk} for different antibody  concentrations $c=50, 100$ and $200$ mg/ml and for the two ionic strengths $I$ under investigation in this work. These are the equivalent, experimentally, to $I=7$ and $57$ mM, the latter being obtained by adding salt to the solution. In the following, we will refer to these two ionic strengths either with explicit reference to the experimental conditions or by simply mentioning a solution with or without added salt. 
The major difference between the potentials ascribed to the two ionic strengths lies in the presence of a repulsive tail for $I=7$ mM.
%In fact, the lack of additional salt in the solution causes interactions between antibodies not to be screened at all concentrations except for $c=300$ mg/ml, that is the highest we investigate. 
On the contrary, by increasing the salt concentration, electrostatic interactions are screened and the short-range attraction term prevails, leaving limited differences in the depth of the attractive well for varying concentrations. Note that the explicit concentration dependence of the potential arises from the contribution of the counterions to the screening length, and also incorporates the Gibbs-Donnan effect encountered in the preparation of the high concentration samples~\cite{gulotta2024combining}.

We then perform equilibrium molecular dynamics (MD) simulations of coarse-grained antibodies in an implicit solvent for different concentrations and ionic strengths
%, as from the potentials in Figure~\ref{fig:potentials_yuk} 
(see Methods). 
By leveraging the antibody equilibrium trajectories, we also have access to relevant information on their dynamical properties. In fact, the integral of the stress autocorrelation function $C_{\sigma\sigma}(t)$ is directly linked via a Green-Kubo relation to the zero-shear viscosity $\eta$, according to Equation~\ref{eq:etagreenkubo} (see Methods).
%More details on the calculation of $C_{\sigma\sigma}(t)$ are given in Methods.

The stress autocorrelation function $C_{\sigma\sigma}(t)$ as a function of simulation time $t/\tau$, with $\tau$ the unit of time in simulations, is shown in Figure~\ref{fig:stress_visc_yuk}(a.I) for three of the five investigated concentrations and for the two ionic strengths. The correlation function for the lowest and highest concentrations is reported in Figure S5. Figure~\ref{fig:stress_visc_yuk}(a.II) highlights the long-time behavior of $C_{\sigma\sigma}(t)$, where it  tends to zero. 
%The corresponding relative viscosity $\eta_{rel}$ is reported in Figure~\ref{fig:stress_visc_yuk}(c).
We note that at short times, the correlation is always larger for the highest ionic strength $I=57$ mM as compared to the case in which no salt is added to the solution. However, for $t \gtrsim \tau$, the behavior of the autocorrelation function is inverted, with the highest ionic strength correlation function reaching zero faster than the lowest ionic strength (see Figure~\ref{fig:stress_visc_yuk}(a.II)). 

The shape of $C_{\sigma\sigma}(t)$ clearly affects the value of the viscosity $\eta$, which is reported in Figure~\ref{fig:stress_visc_yuk}(b). As expected, the viscosity of the system increases with concentration. However, we also observe viscosity values for the lowest ionic strength that are consistently higher, across the whole range of studied concentrations, than those for $I=57$ mM. Remarkably, this behavior is in contrast to what observed in experiments~\cite{gulotta2024combining} for the same antibody, where a similar relative viscosity for the two ionic strengths was measured up to $c \leq 100$ mg/ml but a net increase was found for higher concentrations for $I=57$ mM.
%difference was found for higher concentrations, with the $I=57$ mM system viscosity being significantly higher than that of the lower ionic strength. 
The reason for this finding in the simulations lies essentially in the inversion of the correlation function at long times, whose weight in the integral in Equation~\ref{eq:etagreenkubo} is greater than for short times, thus yielding a larger viscosity for the low ionic strength.
From this analysis, we thus find that a simple model  based on effective Yukawa interactions and uniform charge distribution fails to appropriately describe the collective rheological behavior of concentrated solutions of antibodies in the presence of different salt concentrations, despite being able to correctly reproduce their equilibrium structural behavior.
\\
\\
\indent To improve this description, we thus move to a more accurate modeling of the antibodies where their heterogeneous charge distribution is taken into account. In fact, while the overall effective charge is positive, on a smaller scale there exist differences within areas of its domains, as clearly visible from the electrostatic iso-potential surfaces shown in Figure S2. 
To make the model as realistic as possible, we map the charges directly from the amino acid representation of the antibody onto the coarse-grained bead-based model. This procedure is schematically outlined in Figure~\ref{fig:models}(d). At first, we run an all-atom molecular dynamics simulation of a single antibody molecule to allow for the relaxation of the structure. Subsequently, we perform a first coarse-graining by grouping together in single beads atoms that belong to the same amino acid. For assigning the charge on each amino acid, 
%we equilibrate the structure by performing titration moves at the experimentally desired conditions. 
we perform a one-protein Monte Carlo constant pH simulation at the experimentally desired conditions. 
In this way, we obtain the model reported in Figure~\ref{fig:models}(d.I), where the color coding reflects the charge value assigned to each bead. After appropriately rescaling the amino acid onto the bead model, we are then able to assign the respective charges by determining the closest distance to each of the beads (see Figure~\ref{fig:models}(d.II)). More details on the modeling can be found in Methods.
The model we now adopt is depicted in Figure~\ref{fig:models}(b).
%, while in Methods we describe how it is obtained.
We thus repeat simulations with this model as for the one with a uniform charge distribution. In particular, we first verify the consistency of the static structure factors with experiments (see Figure S4) and then we calculate the stress autocorrelation function $C_{\sigma\sigma}(t)$ for different concentrations and ionic strengths. The latter is reported in Figure~\ref{fig:stress_visc_yuk}(c.I), while in the respective panel II we show an enlargement of the long-time contribution. The same quantities for the lowest and highest concentrations are shown in Figure S6. Despite the overall trends as a function of concentration and ionic strengths being confirmed, we immediately note that the decay of the correlation function occurs for this model on a different, much shorter, time scale than for the uniformly charged case.  
Notwithstanding this, the long-time correlation displays similar features to the ones emerging from a uniform charge distribution, as it can be seen by comparing Figure~\ref{fig:stress_visc_yuk}(a.II) and (c.II). In particular, also the zero-shear viscosity calculated for this model, reported in Figure~\ref{fig:stress_visc_yuk}(d), is still higher for $I=7$ mM than for $I=57$ mM, albeit the differences between the two curves being narrower with respect to the uniformly charged case.

\subsection*{Dissecting the different contributions to the viscosity}
\indent For better understanding the origin of the different contributions to the viscosity, we analyze the stress autocorrelation function arising from additional simulations in which the bead-bead interactions are specifically designed to either exclude, or include solely, the short-range attractive term of the overall potential (see Methods). The results reported in Figure~\ref{fig:attr_noattr_yuk} show the stress autocorrelation function $C_{\sigma\sigma}(t)$, and the same quantity normalized by $C_{\sigma\sigma}(0)$, for both implicit ions models with uniform and non-uniform charge distributions for the representative concentration $c=200$ mg/ml and ionic strength $I=7$ mM. 
%For the case with uniform charges, 
The bead-bead potentials specifically conceived for this analysis are shown for the case with like-charges as inset in Figure~\ref{fig:potentials_yuk}. %, together with the full potential reproducing the structure factor of the laboratory system. 
Both for uniform and heterogeneous charge distributions, a fully attractive potential favors the increase of the stress correlation at short times, see empty symbols in Figure~\ref{fig:attr_noattr_yuk}(a,b.I), whereas a decay at longer times is observed for a potential with predominantly long-range repulsive interactions, see full symbols in Figure~\ref{fig:attr_noattr_yuk}. With respect to our case of interest, this explains why for the highest ionic strength we always observe a higher correlation at short times, being the corresponding potential essentially constituted by an attractive well, see Figure~\ref{fig:potentials_yuk}. On the contrary, the presence of a repulsive tail in the potential for the $I=7$ mM case is responsible for the much higher viscosity at low ionic strength. Here, the contributions to the integral of the stress autocorrelation function will be always higher than for the case in which charges are screened.
\\
Important insights also emerge by treating two  distinct charge distributions. In fact, the inclusion of charge inhomogeneities, which also  favors interactions between beads of opposite sign, brings along two relevant consequences in the shape of the stress autocorrelation function. 
%First, we note that, when the beads of the antibody have different charges, the characteristic time of the decay is shorter by roughly two orders of magnitude. 
First, we note that, when there are beads with different and opposite charges, the characteristic decay time decreases by approximately two orders of magnitude.
%In addition, the decay shows a pronounced secondary relaxation at intermediate times, as compared to the uniform-charge case for which $C_{\sigma\sigma}(t)$ decays smoothly to zero at longer times, see Figure~\ref{fig:attr_noattr_yuk}(a,b.II).
In addition, the secondary relaxation that we observe at intermediate times is much more pronounced as compared to that found for the uniform-charge case, where it appears that $C_{\sigma\sigma}(t)$ decays smoothly to zero at longer times, see Figure~\ref{fig:attr_noattr_yuk}(a,b.II).
%The presence of an additional relaxation mode, which is also found in the full potential treatment (see Fig.~\ref{fig:stress_visc_yuk}), is even more evident in the case where only Yukawa interactions are present, since the inclusion of a short-range attraction has the effect of smearing out and shortening the long-time decay of the correlation.
This feature which is also found in the full potential treatment (see Figure~\ref{fig:stress_visc_yuk}), is even more evident in the case where only Yukawa interactions are present, since the inclusion of a short-range attraction has the effect of smearing out and shortening the long-time decay of the correlation.
Regardless of whether the integral of $C_{\sigma\sigma}(t)$ gives rise to  different absolute values, this analysis suggests that the mechanism underlying the increase in viscosity has different origins for the two cases. In the uniform charge case, it is reasonable to hypothesize that the main contribution to viscosity is simply given by the increase in concentration of a rather stable solution of antibodies. On the contrary, the presence of differently charged domains allows for a faster decorrelation of the inherent stress, although this happens in two distinct steps, short- vs intermediate-times. 
%The latter may be tentatively attributed to a transient electrostatic attraction between oppositely-charged regions of the antibody. 
%However, it is evident that the characteristic time of this secondary relaxation is too short to be compared with experimental observations.

\subsection*{Accounting explicitly for charges, counterions and salt ions}

\begin{figure*}[t!]
\centering
\includegraphics[width=0.95\textwidth]{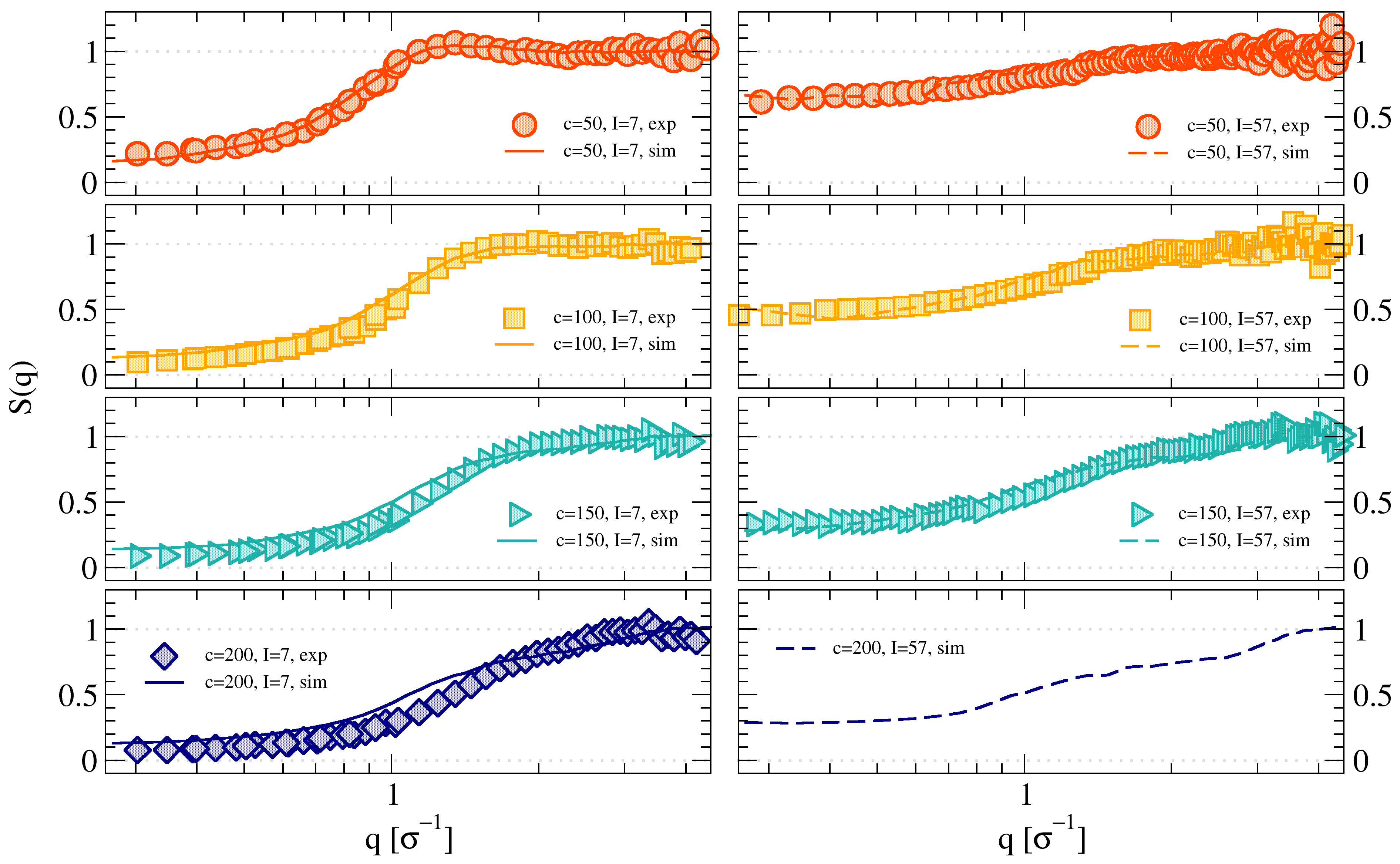}
\caption{\small \textbf{Explicit ions model static structure factor.} Static structure factor $S(q)$ as a function of the scattering vectors $q$ for ionic strengths (left) $I=7$ mM and (right) $I=57$ mM for increasing antibody concentrations $c$ (from top to bottom), as from the legends, for simulations (full and dashed lines, respectively) and experiments (symbols). Data are for simulations with explicit charged antibodies, counterions and salt.
Grey dotted lines are guides-to-the-eye for $S(q)=0$ and $1$. The experimental data for $c=200$ mg/ml and $I=57$ mM are not available.
}
\label{fig:sq_exp}
\end{figure*}

The previous analysis suggests that in order to properly capture the behavior of the viscosity, it is necessary to adopt a more accurate model allowing us to reproduce correctly long-time stress correlations, beside the non-uniformity of the charge distribution. We thus move to a modeling based on explicit Coulomb interactions between charges, considering ions and counterions that might mediate antibody charge-charge interactions in a more accurate way than by means of fully implicit screened potentials. 
We thus design a new primitive model in which we retain the anisotropic shape of the antibodies and assign explicit charges to each bead by keeping a heterogeneous distribution as for the last model analyzed.
%. The charge distribution is similar to the one adopted for the implicit ions model with heterogeneous charges.
Concerning interaction potentials, we keep the same excluded volume and short-range attractive  interactions as for the implicit ions model, only slightly calibrating the level of attraction at short distances in order to match the experimental structure factors. Other details on this model are reported in Methods.

Before exploring the rheological behavior of the antibodies, we need to verify whether the explicit ions modeling correctly reproduces the equilibrium behavior. To this aim, we prepare initial configurations by adding monovalent counterions and ensuring system neutrality, thus reproducing the $I=7$ mM condition. Importantly, for studying the highest ionic strength, that is $I=57$ mM, we include an appropriate amount of oppositely-charged salt ions to mimic the $50$ mM added salt concentration in experiments (see Methods).
The static structure factors $S(q)$ as a function of the wavenumber $q$ for our primitive model are shown in Figure~\ref{fig:sq_exp} for four representative antibody concentrations, namely $c=50, 100, 150$ and $200$ mg/ml, and for the two studied ionic strengths $I=7$ and $57$ mM. The experimental structure factors previously measured by some of us~\cite{gulotta2024combining} are also reported in Figure~\ref{fig:sq_exp} for the same conditions.
The agreement between numerical and experimental structure factors is remarkable at all investigated conditions, with little deviations only at intermediate wavenumbers for $c=200$ mg/ml, similarly to what formerly reported for the Yukawa case~\cite{polimeni2024multi}. In particular, the effect of salt addition, which screens the repulsion and causes attraction to be felt more strongly, is also well reproduced. Indeed, the structure factors increase at low scattering vectors as compared to the case where no salt is added. This implies that the macroscopic behavior of the system is adequately considered by our modeling and that the interaction potentials arising by having explicit charges well reproduce the molecular peculiarities of the antibodies. Besides, it correctly captures the antibody concentration and salt effects in the complex electrostatics of the system.

\begin{figure*}[t!]
\centering
\includegraphics[width=\textwidth]{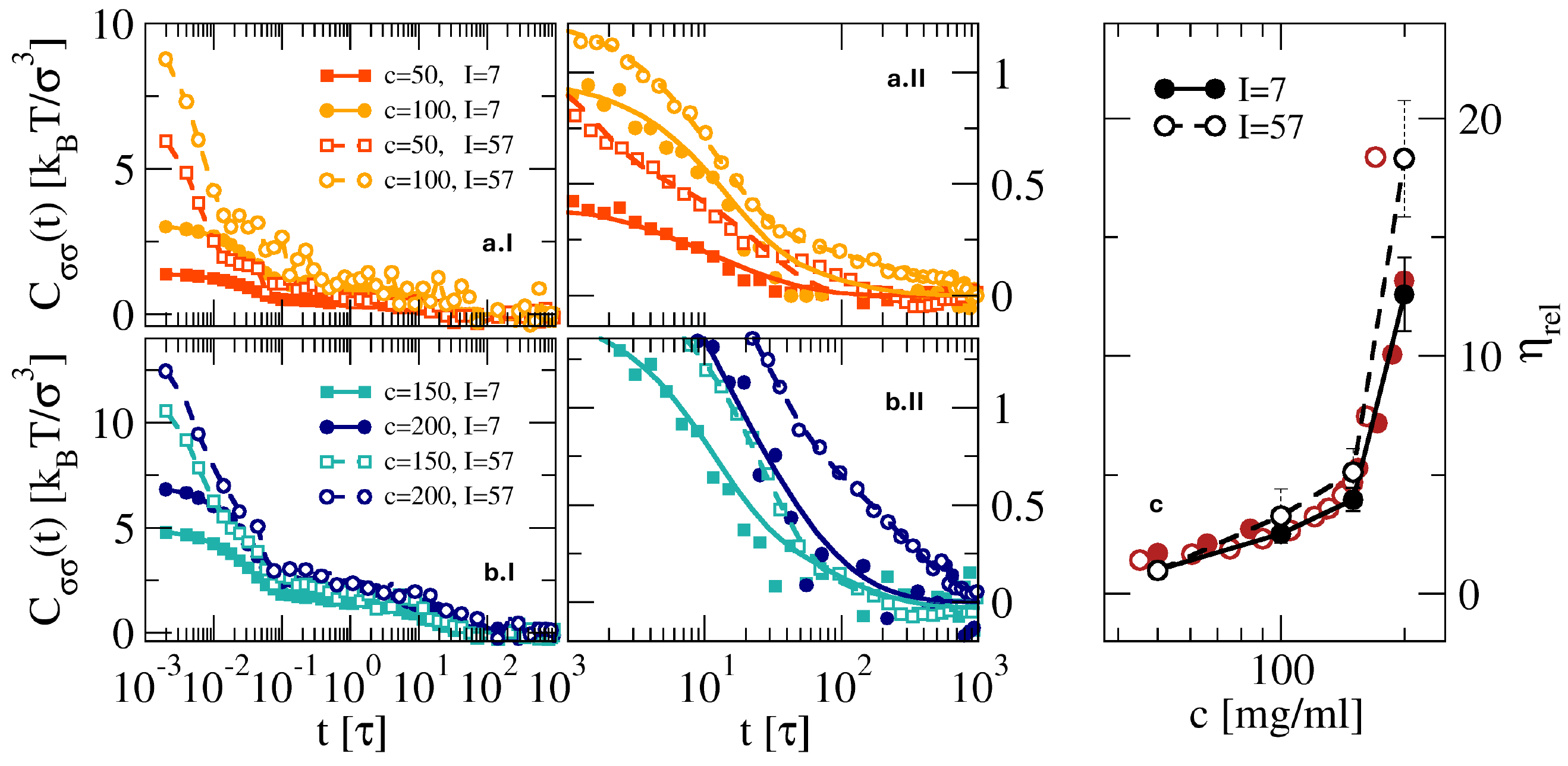}
\caption{\small \textbf{Rheological response for the explicit ions model.} (a,b) Stress autocorrelation function $C_{\sigma\sigma}(t)$ as a function of simulation time $t$ for ionic strengths $I=7$ mM (full lines) and $I=57$ mM (dashed lines) for different antibody concentrations $c$ (reported in mg/ml), as from the legend. Panels II show an enlargement of $C_{\sigma\sigma}(t)$ for the long-time correlations. Full and dashed lines are guides-to-the-eye for $I=7$ mM and $I=57$ mM, respectively. (c) Relative viscosity $\eta_{rel}$ from experiments (red symbols)~\cite{gulotta2024combining}, as a function of antibody concentration $c$ and from simulations (black symbols) with explicit charged antibodies, counterions and salt. The latter are rescaled on the experimental data for ionic strengths $I=7$ mM (full line and filled symbols) and $I=57$ mM (dashed line and empty symbols) and reported with relative error bars, as detailed in Methods.
}
\label{fig:stress_visc_exp}
\end{figure*}

We then analyze the rheological behavior that emerges from the explicit ions modeling. We make use of the same trajectories obtained for calculating the static structure factors and determine the stress autocorrelation function and the viscosity of the system, exploiting the same protocol used for the implicit ions models.
The obtained $C_{\sigma\sigma}(t)$ are shown in Figure~\ref{fig:stress_visc_exp}(a,b) for different concentrations and ionic strengths.
%, whereby panel c reports the zero-shear viscosity.
The decay of the stress correlation at short times is comparable to that of the implicit ions modeling, with the highest ionic strength having the largest signal. However, the use of explicit ions introduces striking additional features. Besides the substantial increase in the value of $C_{\sigma\sigma}$, it presents a significantly marked secondary relaxation of the stress at intermediate times for all investigated conditions, extending to much longer times (almost two decades) than for the implicit charged models. This feature is reminiscent of the behavior observed for the implicit model with heterogeneous charge, although it is now much more pronounced. This points to a similar mechanism of the relaxation of the stress, given by the presence of different charges on the beads of the antibodies. Most importantly, the relaxation functions display a decay that is faster for the lower ionic strength, see Figure~\ref{fig:stress_visc_exp}(a,b.II). This trend is consistent for the entire range of investigated concentrations, with the stress always relaxing at slightly longer times for the case with added salt.

\begin{figure*}[t!]
\centering
\includegraphics[width=\textwidth]{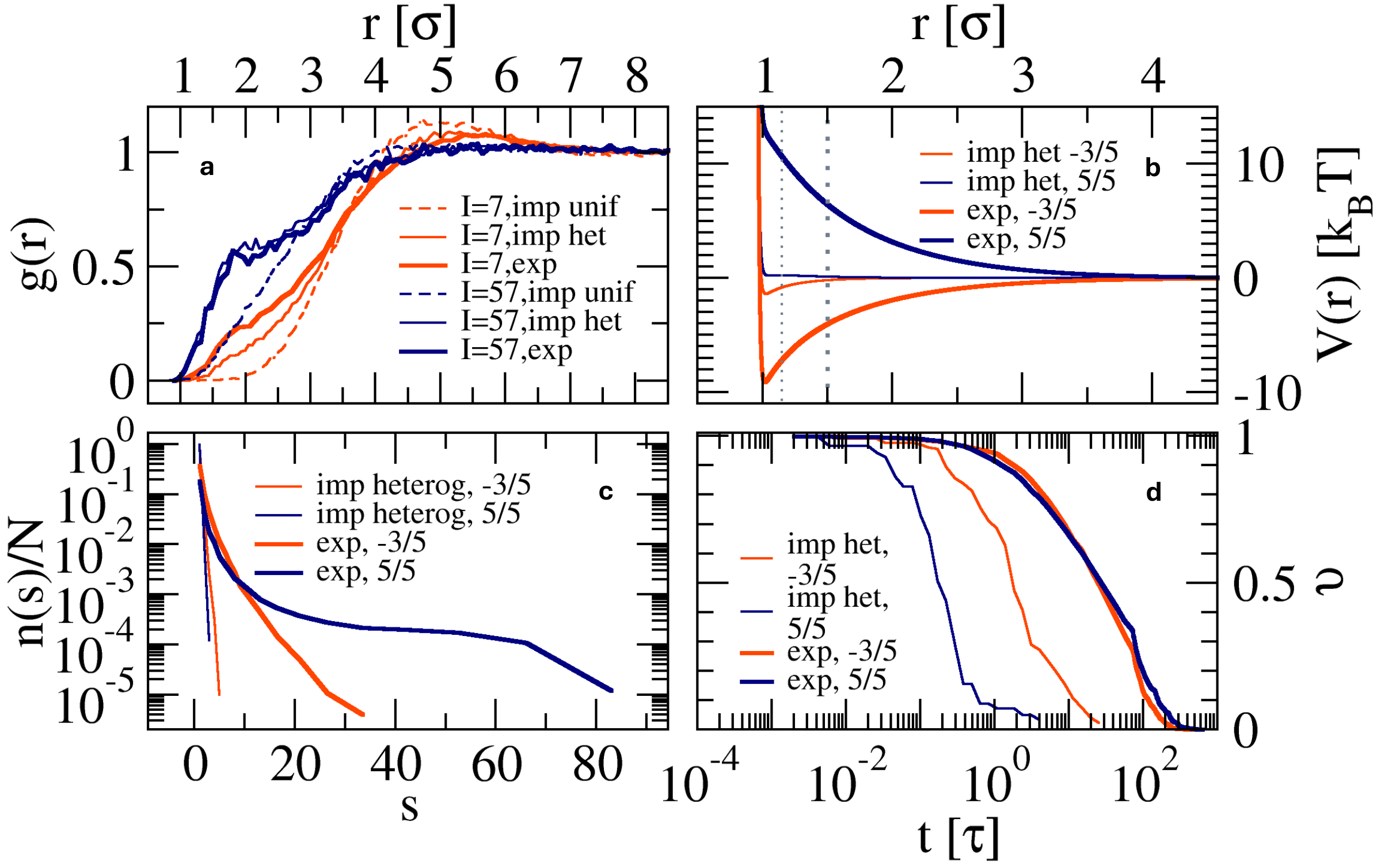}
\caption{\small \textbf{Transient interacting antibodies.}
(a) Antibody center-of-mass radial distribution function $g(r)$ as a function of the distance $r$ for $c=200$ mg/ml for two ionic strengths $I=7, 57$ mM for all the models investigated. (b) Bead-bead interaction potential $V(r)$ as a function of the distance $r$, (c) size distribution $n(s)$ of transient aggregates of antibodies of size $s$, (d) environment correlation function $\upsilon$ as a function of simulations time $t$ for the implicit ions model with heterogeneous charge distribution (thin lines) and for the explicit ions model (thick lines), for the interaction between beads with charges $-3/5$ (orange lines) and $5/5$ (blue lines), for a concentration $c=200$ mg/ml and ionic strength $I=57$ mM. The thin and thick dotted lines in (b) indicate the distance at which $V(r)=1$ and $4k_BT$.
}
\label{fig:clusters_exp}
\end{figure*}

%The corresponding viscosity $\eta$ is reported in Figure~\ref{fig:stress_visc_exp}(c), showing that the system at the highest ionic strength displays an enhanced viscosity as compared to that at a low ionic strength. The absolute values of $\eta$ are significantly higher than for the implicit ions modeling, as a result of the fact that $C_{\sigma\sigma}(t)$ relaxes within almost six decades in time.
The corresponding viscosity is reported in Figure~\ref{fig:stress_visc_exp}(c). Here we show the numerically-calculated viscosity $\eta$ rescaled on the experimental viscosity measured for this system~\cite{gulotta2024combining}. For this reason, we directly report the relative viscosity $\eta_{rel}$, as measured in experiments (see Methods). The absolute values of $\eta$, shown in Figure S7, are significantly higher than for the implicit ions modeling, since $C_{\sigma\sigma}(t)$ relaxes within almost six decades in time.
We find that the system at the highest ionic strength displays an enhanced viscosity as compared to that at a low ionic strength. We note how the viscosity steadily increases for $c \leq 150$ mg/ml with only little differences between the two ionic strengths. However, it presents a sharper growth at higher concentrations, with $\eta$ that increases by a factor of $\approx 4$ for $I=57$ and by a factor of $\approx 3$ for the case in which no salt is added. Notably, this picture is found to be in qualitative agreement with experimental results~\cite{gulotta2024combining}, in stark contrast to the outcomes obtained with the implicit charge models.  
It is therefore evident that the explicit treatment of charges succeeds in accounting long-time correlations of the stress in the system in a more realistic manner than for the implicit treatment, leading to a correct qualitative description of the collective dynamics.

\begin{figure*}[t!]
\centering
\includegraphics[width=\textwidth]{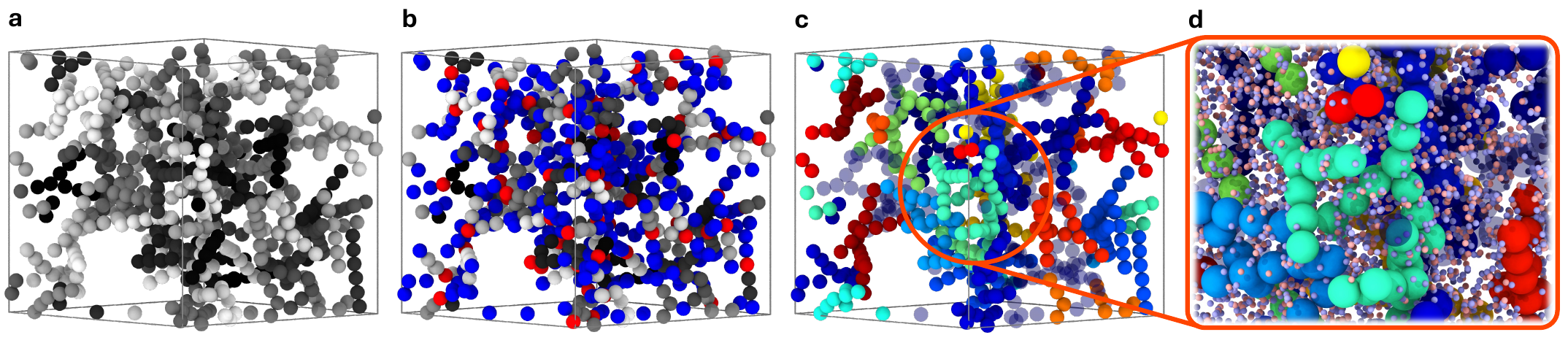}
\caption{\small \textbf{Strongly correlated antibodies.} 
Representative simulations snapshots displaying one of the biggest transient aggregates ($s=74$) formed between $+5/+5$ charges for $c=200$ mg/ml and $I=57$ mM. Specifically, in (a) different antibodies have different shades of gray, in (b) $-3$ and $+5$ charges are colored in red and blue, respectively, and in (c) different clusters between $-3/+5$ have different colors and individual antibodies have a higher transparency. Ions and counterions are not shown for visual clarity. Panel (d) shows an enlargement of panel (c) with the additional representation of ions and counterions as smaller (not-in-scale) light red and blue beads. 
}
\label{fig:snapscluster}
\end{figure*}

Having determined the most appropriate model for describing the viscosity of the system, we now aim to gain a deeper understanding of the macroscopic behavior by analyzing the microscopic structure formed by the antibodies. 
Differences that emerge between models may be indicative of how antibodies arrange in the experimental sample. We recall that the charge distribution on the antibody considered here is such that there are no evident patch regions that could result in equilibrium cluster formation and strongly enhanced viscosity as reported for other antibodies~\cite{skar2019colloid}. 
To monitor structural correlations between antibodies, we report the center-of-mass radial distribution function $g(r)$ in Figure~\ref{fig:clusters_exp}(a) for $I=7$ mM and $57$ mM, and $c=200$ mg/ml, for all the models investigated. The model with uniform charges shows a radial distribution function with no peculiar features, typical of a fluid. This finding corroborates the hypothesis according to which the mechanism for releasing the stress for the uniform charge distribution case is simply linked to objects interacting more closely (and strongly) following an increase in concentration. Instead, for both models with heterogeneous charge distribution, we note the emergence of a similar shoulder at $2 < r/\sigma < 3$, at distances that are slightly greater than the radius of gyration of the antibodies. This peak becomes particularly evident for the highest ionic strength and accounts for the presence of antibody pairs that are closer to each other.

For investigating the microscopic arrangement of antibodies, we first analyze the interactions at play between beads of the antibodies. Doing so, we aim to define, for each model, a characteristic distance within which antibodies have significant interactions, at least exceeding the thermal energy in magnitude. 
We focus on the case $c=200 $ mg/ml at the highest ionic strength, since it gives rise to a higher peak in the $g(r)$. The same analysis is reported in Figure S9-S10 for a lower concentration and ionic strength, showing similar trends as the ones described hereafter.
In Figure~\ref{fig:clusters_exp}(b) we report the bead-bead interaction potential for the implicit and explicit ions models considering the charge pairs that allow for the highest electrostatic attraction and repulsion, namely $-3$ and $+5$, and $+5$ and $+5$ pairs, respectively. It is clear that charged Coulomb interactions are significantly stronger than for the implicit case, for which the minimum energy is just comparable to the thermal energy. Therefore, while in the implicit ions model we can only choose a distance at which the oppositely-charged potential is around $-1k_BT$, in the case of the explicit ions model, we consider a distance at which the $-3/+5$ interactions reach at least $-4k_BT$.
%, which should give rise to longer-lived interactions. 
For simplicity, we retain the same distance also for $+5/+5$ interactions, where the respective repulsive  interaction potentials have a comparable magnitude. We report in Figure~\ref{fig:clusters_exp}(c) the size distribution $n(s)$ of the antibodies interacting within the respective distances normalized by the number of antibodies $N$, and in Figure~\ref{fig:clusters_exp}(d) the \textit{environment} autocorrelation function $\upsilon (t)$ of such interactions. This is calculated as the autocorrelation between the number of interacting antibodies within the chosen distance, $N_\upsilon(t)$ at time $t$, and the same quantity in an initial configuration $N_\upsilon(0)$ (see Methods). %$\upsilon (t)$ is analogous to what sometimes considered in repulsive glass systems\cite{zaccarelli2009} and named `cage' autocorrelation function.
$\upsilon (t)$ is analogous to what sometimes is referred to as cage autocorrelation function in repulsive glass systems~\cite{zaccarelli2009colloidal}.
Here, since there is no real cage, we define this as a correlation function of the environment around the antibodies. 
%For oppositely charged beads, the interactions between neighboring antibodies can be regarded as bonds. Consequently, these antibodies can be considered to form clusters in the conventional sense. In contrast, this does not apply to strongly repulsive like-charges, which instead form strongly interacting, correlated pairs or groups of antibodies.
For oppositely charged beads, it is tempting to view the attractive interactions between the beads as temporary bonds between different antibodies. Consequently, these antibodies could then be considered  to form transient clusters in the conventional sense. In contrast, this does not apply to strongly repulsive like-charged pairs of beads, which can be just thought of forming strongly interacting and highly correlated pairs or groups of antibodies.
%instead indicate the formation of strongly interacting and highly correlated pairs or groups of antibodies.

As we can expect just by looking at the involved potentials, in the system with effective screened Coulomb interactions the linked antibodies are few, they form small groups, and their encounters are very short-lived, leaving more than $90\%$ of the antibodies to behave as individual objects. While thermal fluctuations might bring them temporarily at short distances, they do not experience a force that is strong enough to hold them close. Under these conditions, the increase of the viscosity with increasing concentration is thus driven primarily by the excluded volume of the antibodies, and it is almost independent of ionic strength.
On the contrary, in the system where charges and ions are treated in an explicit manner, proximate interactions between beads involve a wide number of antibodies, since they are even able to span the entire simulation box. As an example, a representative simulation snapshot where a large number of antibodies ($s=74$) is involved with $+5/+5$ interactions is illustrated in Figure~\ref{fig:snapscluster}(a,b) with beads belonging to the same molecule colored with the same shade of gray, and with $-3$ and $+5$ charges depicted in red and blue, respectively. In panel c, we further show that within the very same structure also interactions between beads with opposite charges are present. Moreover, we highlight the important role played by explicit counterions and salt ions by showing them in panel d.
It is furthermore highly instructive to look at the temporal evolution of $\upsilon(t)$ for both models. For the implicit ions model, we observe that the environment around $-3/+5$ pairs decorrelates at least an order of magnitude more slowly than that of $+5/+5$ pairs. This is in line with a qualitative common expectation that close proximity between repulsive pairs of beads would be less favorable and thus more short lived than for attractive pairs. The observed behavior changes completely in the case of the explicit ions model. Not only does $\upsilon(t)$ decay much more slowly than for the implicit ions model, but both types of pairs have comparable life times. Moreover, $\upsilon(t)$ decays within a time window that is comparable to the final decay of the stress autocorrelation function. This indicates that it is the decay of these strongly correlated transient structures that determines the zero shear viscosity, somewhat analogous to the link between the relaxation of the nearest neighbor cage structure and the viscosity for hard sphere colloids. While the flow behavior of the system is thus linked to the build up and subsequent decay of transient structures at high concentrations, these correlated structures are different from the commonly proposed transient clusters formed by oppositely charged domains. Based on our study, these only form when explicitly treating all macroion charges and small ions, and not relying on an effective pair potential based on a screened Coulomb interactions.

\section*{Discussion}

One of the key reasons for the substantial interest in investigating the rheological properties of antibody solutions is that they have direct implications for the administration of drugs that are effective, biologically, for different purposes. From a fundamental perspective, these solutions are constituted by anisotropic, charged objects, thus being one of the favorite model systems in the realm of colloidal physics. It is thus tempting to use a colloid physics-based approach in order to obtain a quantitative and predictive understanding of the flow properties of concentrated antibody solutions.
For achieving this, there are two interrelated aspects that need to be considered. The first is the development of models that accurately reflect reality and replicate experimental evidence. Once this is accomplished, the subsequent step is to extract information to comprehend the underlying microscopic causes of the observed behavior at the macroscopic level. 
%To this end, we initiated our investigation with one of the most simple models that is commonly utilized in this context. Specifically, we used an anisotropic coarse-grained bead-based model where charged interactions are treated with implicit ions and beads all have the same charge. This is compared with another model in which the only difference is that charge is distributed non-uniformly on the beads. In this way, we were able to reproduce more closely the charge distribution that real IgG1 antibodies have. While both these models reproduce correctly the structure of experimental systems for different concentrations and ionic strengths, they turn out not to be suitable for describing the viscosity of our system. For the first time, this is not obtained by indirect extrapolation or fits, but by directly calculating the stress auto-correlation function between antibodies and relating it to the viscosity via the fundamental Green-Kubo relation. 
%Given this, perhaps unexpected, failure, the same approach was thus leveraged for verifying that a model in which charges interact via a long-range Coulomb potential is capable of describing both the structure and the viscosity of the system. In fact, only by introducing explicit charges and related counterions and salt ions, we were able to reproduce experimental evidence, showing that addition of salt in solutions causes the viscosity to be higher than the case in which only antibodies are dispersed in the buffer.

To this end, we initiated our investigation using computer simulations based on simple and previously established coarse-grained and bead-based models of antibodies. 
Viscosities for these systems were determined by directly calculating the integral of the stress autocorrelation function, as prescribed by the fundamental Green-Kubo relation, rather than relying on indirect extrapolation or phenomenological theoretical models.
We first started with an anisotropic model where charge interactions are treated with implicit ions and where all beads have the same charge. This model was subsequently modified such that the charge is distributed non-uniformly between the beads to reflect more closely the actual charge distribution on the real antibody under investigation.
While both models accurately reproduce the solution structure previously determined through small-angle X-ray scattering experiments as a function of concentration and ionic strength, they were clearly unsuccessful in describing the viscosity of these systems.
In response to this (perhaps unexpected) failure, we moved beyond the traditional use of an effective potential, where counterions and salt ions are included only implicitly through a screening parameter, and instead introduced explicit charges and ions interacting through a long-ranged Coulomb potential.
This model is indeed capable of describing both the structure and the viscosity of the system. In fact, only in this way we were able to reproduce experimental evidence, showing that for the antibody considered here the addition of salt in solutions causes the viscosity to be higher than for the case in which only antibodies are dispersed in the buffer.

Thanks to our detailed numerical investigation, we could thus unveil several important aspects that the widely-used Yukawa model with implicit charges fails to capture. First of all, we note how the uniformly charged model has a stress relaxation mechanism that distinguishes it from the other two models analyzed. The behavior, based on both the decay of the stress correlation function and the spatial distribution of antibodies, is nothing more than that of a system in which the repulsive interactions between objects get stronger as a result of the proximity that a higher concentration implies. This is similar to what has previously been established for charged colloids, where the relative viscosity of the suspension strongly increases with decreasing ionic strength for a given concentration~\cite{Horn2000}. Therefore, the first relevant point that can be drawn from this analysis is that there is the strong need to preserve the realistic, heterogeneous charge distribution of individual antibodies. 

% However, our investigation also shows that this is not enough for reproducing the experimental findings when using an implicit ions model. Here, the magnitude and the nature of interactions involved are not sufficient for generating a decay of the stress at long times that gives rise to the observed higher viscosity upon increasing salt concentration.
% In fact, while attractive interactions between oppositely charged Yukawa beads successfully capture the experimentally observed equilibrium solution structure and induce a weak two-step relaxation in the stress autocorrelation function, the resulting zero-shear viscosity does not reflect the experimental evidence.
% It is only when using a heterogeneous charge distribution combined with explicit counter- and co-ions and Coulomb interactions between charged species that we find long-range stress correlations within a much more extended time window, exceeding those of the Yukawa model by roughly two decades in time.
However, our investigation reveals that an implicit ions model is insufficient to reproduce the experimental findings. Specifically, the magnitude and nature of the interactions in this approach fail to generate the long-time stress decay required to explain the observed increase in viscosity with higher salt concentrations.
While attractive interactions between oppositely charged Yukawa beads successfully capture the experimentally observed equilibrium solution structure and induce a weak two-step relaxation in the stress autocorrelation function, the resulting zero-shear viscosity remains inconsistent with experimental evidence.
Only by employing a heterogeneous charge distribution combined with explicit counter- and co-ions, along with Coulomb interactions between charged species, we do  observe long-range stress correlations that persist over a much broader time window, approximately two orders of magnitude longer than those predicted by the Yukawa model.
%The explicit ions model qualitatively captures both the measured concentration dependence of the viscosity and its growth with increasing ionic strength.
%However, our investigation also shows that the magnitude and nature of interactions described by screened Coulomb potentials are insufficient to produce the stress decay required to explain the observed increase in viscosity with added salt. Only when incorporating a heterogeneous charge distribution, explicit ions, and Coulomb interactions between charged species we do observe long-range stress correlations that persist over a much longer time window, exceeding the Yukawa model by approximately two decades in time.

Increased solution viscosity in highly concentrated antibody solutions was previously often related to the formation of transient clusters.
In the present work, the center-of-mass radial distribution functions of antibodies for the explicit ions model in Figure~\ref{fig:clusters_exp}c, with a shoulder just beyond the antibody radius, may indeed suggest the formation of clusters. This could lead to the assumption that oppositely charged beads serve as transient bonds within these structures, and that the larger clusters observed at high ionic strength are responsible for the ionic strength dependence of $\eta$. However, our detailed analysis of the interaction contributions to the environment autocorrelation function $\upsilon(t)$ demonstrates that this interpretation is overly simplistic. While we conclude that the increased viscosity at higher ionic strength arises from the formation of extended correlated structures, our findings indicate that these structures are not solely stabilized by electrostatic attractions between oppositely charged beads. Instead, they rely on the explicit inclusion of macroion charges and small ions to account for the observed behavior.

In summary, the present investigation suggests that the increase in viscosity observed experimentally for this system can only be captured by appropriately treating electrostatic contributions, and therefore by explicitly including counterions and salt ions. This is the crucial ingredient through which an accurate description of the viscosity can be achieved when the concentration is raised above $\approx 150$ mg/ml, which is relevant for applications. 
This further indicates that the mean-field screening assumed in simple Yukawa potentials is not good enough to adequately capture the dynamic stress contribution, and thus the viscosity, associated with long-range Coulomb interactions in this system, which is characterized by shape and charge anisotropy.
Clearly, this phenomenon will need to be further investigated for different antibodies, where the balance between attractive and repulsive contributions is different than in the present case.
For antibodies undergoing clustering~\cite{skar2019colloid,chowdhury2023characterizing,skar2023using} and phase separation~\cite{wang2011phase,ausserwoger2023surface,sibanda2023relationship}, a simplified treatment of electrostatics might be sufficient to qualitatively account for the experimentally observed viscosity. However, this hypothesis remains to be tested in future investigations.

This study, therefore, defines the microscopic cause of an enhanced viscosity in the formation of highly correlated transient structures. This insight enables the prospective design of strategies to prevent such occurrence. Intervention can be directed either at the antibody level, by appropriately engineering it, or at the solution conditions, for instance by exploring additives that limit antibody interactions without compromising their biological efficacy.
Most importantly, the study provides a simulation toolbox and clear guidelines for handling similar systems and accessing reliable information about their dynamic properties. Through this approach there is thus no need to rely on low concentration data combined with phenomenological viscosity relationships or overall interaction parameters such as 
%$B_2$ or $k_D$. 
the second virial coefficient or the diffusion interaction parameter.
Beyond antibodies, we envision the application of similar models to other proteins and biological objects.

A final consideration can be made regarding the accuracy of the employed model. 
%We have demonstrated that, qualitatively, the viscosity is well reproduced. 
Moving forward, it will be necessary to investigate which additional factors would enable a more quantitatively accurate comparison. For instance, the role of antibody flexibility at the hinge region, where the three domains are linked, remains largely unexplored in the literature~\cite{stradner2020potential}. It is conceivable that, in highly concentrated systems, individual antibodies might subtly adapt their structure, resulting in qualitatively different interactions. 
Besides, recent advancements in incorporating charge heterogeneity in anisotropic colloids have been proposed, warranting further exploration of their impact on antibodies~\cite{notarmuzi2024features,campos2024machine}. Additionally, evidence of charge regulation phenomena in small biological molecules suggests that these effects may also merit investigation for their influence on collective properties~\cite{lund2013charge,arzensek2015hofmeister}. Exploring these effects could further enhance our understanding of the system dynamic behavior, and refine predictive models for complex biological environments. Ultimately, it enables the use of biologically effective molecules for the treatment of diseases in ways that are increasingly straightforward and accessible to all.

%\subsection*{Supporting Information Appendix (SI)}

% Authors should submit SI as a single separate SI Appendix PDF file, combining all text, figures, tables, movie legends, and SI references. SI will be published as provided by the authors; it will not be edited or composed. Additional details can be found in the \href{https://www.pnas.org/authors/submitting-your-manuscript#manuscript-formatting-guidelines}{PNAS Author Center}. The PNAS Overleaf SI template can be found \href{https://www.overleaf.com/latex/templates/pnas-template-for-supplementary-information/wqfsfqwyjtsd}{here}. Refer to the SI Appendix in the manuscript at an appropriate point in the text. Number supporting figures and tables starting with S1, S2, etc.

% Authors who place detailed materials and methods in an SI Appendix must provide sufficient detail in the main text methods to enable a reader to follow the logic of the procedures and results and also must reference the SI methods. If a paper is fundamentally a study of a new method or technique, then the methods must be described completely in the main text.

%\subsubsection*{SI Datasets}

% Supply .xlsx, .csv, .txt, .rtf, or .pdf files. This file type will be published in raw format and will not be edited or composed.

%\subsubsection*{SI Movies}

% Supply Audio Video Interleave (avi), Quicktime (mov), Windows Media (wmv), animated GIF (gif), or MPEG files. Movie legends should be included in the SI Appendix file. All movies should be submitted at the desired reproduction size and length. Movies should be no more than 10MB in size.

\section*{Materials and Methods}
\small
%\matmethods{%
% \small
% \section*{Methods}
\noindent \textbf{Materials.} The antibody under investigation is an immunoglobulin G of subclass 1 (IgG1) named Actemra or Roactemra. The amino acid sequence of the antibody was obtained from the patent N. US20120301460. The atomistic structure of the antibody (see Figure S1) was built by homology modeling using the Molecular Operating Environment (MOE) software (Chemical Computing Group, Inc.). A more complete description of the homology model process can be found in Refs.~\cite{gulotta2024combining,polimeni2024multi}. A systematic experimental characterization of the main solution properties as a function of concentration and ionic strength has previously been reported in Ref. \cite{gulotta2024combining}, and the experimental data for the static structure factors and the relative viscosity used in the current work were taken from there.

\noindent \textbf{Amino acid representation of the antibody.} From the atomistic model, we build  a coarse-grained representation at an amino acid level of the antibody (see Figure~\ref{fig:models}(d.I)) mapping each amino acid into a single bead centered in the amino acid center of mass. The size of each bead is calculated based on the amino acid molecular weight following the protocol described elsewhere~\cite{mahapatra2021self}.
The charges are assigned to each bead performing single-protein constant pH MC simulations using \textsc{faunus} (v2.9.1 git 3edf85cf)~\cite{stenqvist2013faunus} with titration moves that propagates back and forward the reaction $A \Leftrightarrow HA$, being $A$ and $HA$ the deprotonated and the protonated form of the amino acid, respectively~\cite{johnson1994reactive}. More details on the protocol for the charge calculation can be found in Ref.~\cite{polimeni2024multi}.
%\fc{to be fixed and completed XXX @Marco}

\noindent \textbf{Antibody coarse-grained bead model.} 
%come mappare da conf di marco
%scelta del nove bead che diciamo
The model for monoclonal antibodies is inspired by coarse-grained models of anisotropic colloids. Each antibody comprises nine beads of unit mass $m$ and diameter $\sigma$, which is also taken as the unit of length in simulations, arranged in a Y-shaped molecule. Each of the three branches (or arms) of the molecules consists of three beads positioned along a line, and forms angles of $150^{\circ}$ and $60^{\circ}$ with the others. Its radius of gyration is $\approx 1.7\sigma$. This peculiar Y-shape is designed to account for the structure of laboratory single-molecule antibodies. Besides, it has been shown that a 9-bead model allows to faithfully reproduce static structure factors similarly to less coarse-grained models with more beads, with the advantage of being lighter in the usage of computational resources~\cite{polimeni2024multi}. Each antibody is treated as a rigid body. We also tested models with the same number of beads and different angles among the arms. However, since there appear differences in the static structure factor only in the intermediate range of scattering vectors, we adopt the combination of angles that reproduces more closely the experimental data (see Figure S9). A more systematic and extensive investigation on this aspect of the model is beyond the scope of the present work.

The coarse-grained model constitutes the basic scaffold for simulations with both \textit{implicit} and \textit{explicit} ions (see also below). For having a non-uniform charge distribution, charges assigned to each bead are mapped from the amino acid representation of the antibody, obtained as described in the previous subsection.
%While in the former case, each bead is assigned the same charges as done in Ref.~\cite{polimeni2024multi}, for the latter model they are mapped from the aminoacid representation of the antibody, obtained as described in the previous subsection. 
To this aim, the amino acid and the 9-bead representations are overlapped onto each other by rescaling the size of each amino acidic bead in natural units by the average size of all amino acids, that is $\sigma_{amino}^{avg}=2.79$ nm. In this way, the two representations can be directly compared. Subsequently, for the explicit ions model, each monomer in the 9-bead representation is assigned a charge based on the average charge of the amino acid beads that are at the shortest distance, and rounded to make i) the total charge equal to the sum of the charges in the amino acid representation ($Q_{exp}=31$), and ii) the two upper arms having the same total charge. For implicit ions simulations, each bead is either assigned the same (positive) charge ($q=3.1$), as done in a previous work by some of us~\cite{polimeni2024multi}, or the same charge distribution as for the explicit ions modeling, by rounding the charges in such a way that the sum correspond to the total charges of the uniform model ($Q_{imp}=28$). In this way, implicit ions models with uniform and heterogeneous charge distributions are obtained, respectively.

\noindent \textbf{Interaction potentials.} We run simulations both with implicit and explicit ions. In the former case, the effective interaction between each bead of the Y model is given by a Yukawa (Yuk) potential which reads as
\begin{equation}
V_{Yuk} = \lambda_B q_i q_j \left(\frac{e^{\sigma/2\lambda_D}}{1+\sigma/2\lambda_D}\right)^2 \frac{e^{-r_{ij}/\lambda_D}}{r_{ij}},
\end{equation}
where $\lambda_B=0.257\sigma$ is the Bjerrum length, $q_{i,j}$ is the charge assigned to the $i$-th and $j$-th antibody bead, and $r_{ij}$ is the corresponding distance. The inverse Debye screening length accounts for the net electrostatic effects on the antibody and reads as
\begin{equation}
\frac{1}{\lambda_D}=\left( 4 \pi \lambda_B \left[ \left( \frac{1}{1-\phi}\right) Q \rho_{mAb} + 2\rho_{salt} + 2\rho_{buffer} \right] \right)^{1/2},
\end{equation}
where $\rho_{mAb, salt, buffer}$ are the number densities of antibodies, salt and buffer ions, respectively, $\phi$ is the excluded volume of a single antibody and $\frac{1}{1-\phi}$ accounts for the free accessible volume, with $\phi \equiv \phi_{HS} = \rho_{mAb} \pi \sigma_{HS}^3/6$ the excluded volume of a hard sphere of diameter $\sigma_{HS}=2R_g^{HS}=9.64$ nm~\cite{polimeni2024multi}.

For the explicit ions modeling, all beads of the antibodies, counterions and salt ions interact via a Coulomb (coul) potential
\begin{equation}
V_{coul} = \frac{q_i q_j\sigma}{e^{*2}r_{ij}}\epsilon
\end{equation}
where $q_{i,j}$ is the charge assigned to each object, $e^{*}=(4 \pi \epsilon_0 \epsilon_r \sigma \epsilon)^{1/2}$ is the reduced unit of charge, with $\epsilon_0$ and $\epsilon_r$ the vacuum and relative dielectric constants, and $\epsilon$ is the energy unit. 

In both implicit and explicit ions modeling, we account for excluded volume and van der Waals interactions by letting antibody beads interact via a modified Lennard-Jones (mLJ) potential 
\begin{equation}
V_{\rm mLJ}(r)=
\begin{cases}
C\left[\left(\frac{\sigma}{r}\right)^{96}-\left(\frac{\sigma}{r}\right)^{6}\right] & \text{if $r \le 3.5\sigma$}\\
0 & \text{otherwise,}
\end{cases}
\end{equation}
with $C=0.62$ for simulations with implicit ions model with a uniform charge distribution, $C=0.8$ for simulations with implicit ions model with a heterogeneous charged distribution, and $C=0.9$ for simulations with the explicit ions model. These values are fixed by the favorable comparison of the structure factors calculated for each model with experimental measurements.
For the explicit ions model, salt ions and counterions only experience a repulsive Weeks-Chandler-Anderson (WCA) potential to account for their excluded volume as
\begin{equation}
V_{\rm WCA}(r)=
\begin{cases}
4\epsilon\left[\left(\frac{\sigma}{r}\right)^{12}-\left(\frac{\sigma}{r}\right)^{6}\right] + \epsilon & \text{if $r \le 2^{\frac{1}{6}}\sigma$}\\
0 & \text{otherwise.}
\end{cases}
\end{equation}

For analyzing the effect of individual contributions to the overall potential on the shape of the stress autocorrelation function (see below), for the implicit ions models we also run simulations in which the antibody beads either interact via a sum of $V_{WCA}$ and $V_{Yuk}$ or by means of $V_{mLJ}$ only. In this way, we either exclude, or include solely, the short-range attractive term of the overall potential, respectively.

\noindent \textbf{Simulation details and parameters.} We run equilibrium molecular dynamics simulations in the NVT ensemble, fixing the number of antibodies $N=100$. Simulations are run in a cubic box with periodic boundary conditions in the three dimensions. The volume is adjusted for reproducing different values of experimental antibody concentration $20 \leq c/(\rm mg \, ml^{-1})\leq 300$, as detailed in Ref.~\cite{polimeni2024multi}. For all concentrations and ionic strengths analyzed, we fix $\sigma=2.79$ nm, according to the mapping performed in Ref.~\cite{polimeni2024multi}.

In the case of simulations with explicit charges, we make the system neutral by adding a congruent number of positive and negative monovalent counterions. For adding salt with a concentration of $c_{m,salt}=50$ mM, we determine the number of particles to be added in the simulation box by calculating the molar concentration of the antibodies $c_{m,mAb}$ in solution, considering the antibody molecular weight $M_w=148000$ g/mol. The number of salt particles to be added is thus calculated as $N_{salt}=2N \frac{c_{m,salt}}{c_{m,mAb}}$, half of which is assigned a $+1$ charge and the other half a $-1$ charge. Both counterions and salt ions have a diameter $\sigma_{ions}=0.1\sigma$, as from other coarse-grained simulations of colloidal systems~\cite{del2019numerical}.
As a solver for the long-range Coulomb interactions, for $r_{ij}>8\sigma$, we use the particle-particle particle-mesh (pppm) method~\cite{hockney2021computer}. 

We treat the solvent implicitly and we set the reduced temperature $T^*=k_BT/\epsilon=1.0$, with $k_B$ the Boltzmann constant and $T$ the temperature, by means of a Langevin thermostat. Equilibration runs are carried out for at least $1\times10^6 \delta t$, with $\delta t=0.002\tau$ and $\tau=\sqrt{m \sigma^2/\epsilon}$ the unit of time. A subsequent production run is carried out for at least $1\times10^7 \delta t$. For both implicit and explicit charges, we run simulations from at least two different initial configurations for reducing statistical error. In all cases, we use \textsc{lammps}~\cite{thompson2022lammps}, for running simulations in a fast parallel environment.

\noindent \textbf{Measured quantities.} From the trajectories of our simulations, we calculate both static and dynamic quantities. The static structure factors are a relevant quantity for determining whether  interactions among antibodies are appropriately modeled. Furthermore, they can be easily compared to experiments. The static structure factors are calculated as
\begin{equation}
S(q)=\frac{1}{P(q)}\frac{1}{9N}\left\langle\left(\sum_i^{9N}\sin \boldsymbol{qr}_i\right)^2+\left(\sum_i^{9N}\cos \boldsymbol{qr}_j\right)^2\right\rangle
\end{equation}
with the sums running over all $9N$ beads of the system, $P(q)$ being the form factor of a single antibody model, $\boldsymbol{q}$ the scattering vectors, $\boldsymbol{r}_{i,j}$ the positions of the $i$-th and $j$-th bead, and $\langle \cdots \rangle$ indicating an average over different configurations of the system and over different orientations of the scattering vectors $\boldsymbol{q}$.

For numerically calculating the zero-shear viscosity $\eta$ of the system, we employ the Green-Kubo relation~\cite{kubo1957statistical,kubo1986brownian}
\begin{equation}
    \eta=\int_0^{\infty}dt C_{\sigma\sigma}(t)
    \label{eq:etagreenkubo}
\end{equation}
where $C_{\sigma\sigma}(t)$ is the stress autocorrelation function which reads as~\cite{ramirez2010efficient,gomez2024diffusion}
% \begin{equation}
% \begin{split}
%     C_{\sigma\sigma}(t) & = \frac{V}{5k_BT} \left( \left\langle \sigma_{xy}(t)\sigma_{xy}(0) \right\rangle + \left\langle \sigma_{yz}(t)\sigma_{yz}(0) \right\rangle\\ 
%     & + \left\langle \sigma_{xz}(t)\sigma_{xz}(0) \right\rangle) \\
%     & + \frac{V}{30k_BT} \left( \left\langle N_{xy}(t)N_{xy}(0) \right\rangle + \left\langle N_{yz}(t)N_{yz}(0) \right\rangle\\ 
%     & + \left\langle N_{xz}(t)N_{xz}(0) \right\rangle )
% \end{split}
% \end{equation}
\begin{equation}
\begin{split}
    C_{\sigma\sigma}(t) & = \frac{V}{5k_BT} \big( \langle \sigma_{xy}(t)\sigma_{xy}(0) \rangle + \langle \sigma_{yz}(t)\sigma_{yz}(0) \rangle \\
    & \quad + \langle \sigma_{xz}(t)\sigma_{xz}(0) \rangle \big) \\
    & \quad + \frac{V}{30k_BT} \big( \langle N_{xy}(t)N_{xy}(0) \rangle + \langle N_{yz}(t)N_{yz}(0) \rangle \\
    & \quad + \langle N_{xz}(t)N_{xz}(0) \rangle \big)
\end{split}
\end{equation}
where $\langle \sigma_{\alpha \beta}(t)\sigma_{\alpha \beta}(0) \rangle$ is the autocorrelation of the $\sigma_{\alpha \beta}$ component of the stress tensor at time $t$, with $\alpha, \beta=x,y,z$, and $\langle N_{\alpha\beta}(t)N_{\alpha\beta}(0) \rangle$ the autocorrelation of $N_{\alpha\beta} \equiv \sigma_{\alpha\alpha}-\sigma_{\beta\beta}$. Each element of the stress tensor accounts for kinetic and interaction contributions and, when Coulomb interactions are present, also for $k$-space energy contributions. The integral in equation~\ref{eq:etagreenkubo} is calculated for each state point until $C_{\sigma\sigma}$ starts fluctuating around zero. We remark that we integrate directly raw data since other attempts to integrate fits of the data do not necessarily improve the measurement of $\eta$ (not shown). The error bars in Figures~\ref{fig:stress_visc_yuk} and ~\ref{fig:stress_visc_exp} are calculated from  the estimates of $C_{\sigma \sigma}$ for the different runs of the same state point, and by accounting for the possibility of calculating the integral in Equation~\ref{eq:etagreenkubo} up to a slightly different simulation time.
Also, in the main text, we directly report a normalized viscosity. This is calculated by making use of the diffusion coefficient $D$ extracted from simulations at the lowest investigated concentration, which roughly corresponds to the single-molecule diffusion coefficient $D_0$ (not shown). Assuming the validity of the Stokes-Einstein relation at a low antibody concentration, we normalize the viscosity for the highest ionic strength investigated by the inverse of the ratio between $D_0$ of the two ionic strengths.
Finally, we also compare $\eta$ to the experimentally measured relative viscosity $\eta_{rel}$, taken from Ref.~\cite{gulotta2024combining}. The latter is defined as $\eta_{rel}=\eta_0/\eta_s$, where $\eta_0$ is the zero shear viscosity of the antibody solution and $\eta_s$ the solvent viscosity. The two quantities $\eta$ and $\eta_{rel}$ are directly compared by rescaling the former onto the experimental measure at $c=100$ mg/ml, where differences induced by ionic strengths are still limited.

For the analysis of the transient interacting antibodies, we first calculate the radial distribution function $g(r)$ of the centers of mass of the antibodies. Except for the model with implicit ions and uniform charge distribution, it is possible to identify the presence of a shoulder at short distances, suggesting the presence of antibodies that are closer to each other. For studying such aggregates of antibodies, we focus on the highest oppositely and like-charged interactions, that are $-3/+5$ and $+5+/5$, respectively. We thus define a distance $r_0$ between beads at which interactions exceed thermal energy (see Figure~\ref{fig:clusters_exp}(b)). For the case reported in the text, which is for $c=200$ mg/ml and $I=57$ mM, we take $r_0=1.15\sigma$, corresponding to $V(r)\approx -1k_BT$ for the $-3/+5$ pairs, for the implicit charge model with heterogeneous charge distribution and $r_0=1.5\sigma$, corresponding to $V(r)\approx -4k_BT$ for the $-3/+5$ pairs, for the explicit ions model. For simplicity, we assume the same respective distances also for $+5/+5$ pairs. We then use these bead-bead distances to count the number of antibodies $s$ for which beads are closer than $r_0$, obtaining an average size distribution $n(s)$ over the whole simulation run.
For assessing the transient nature of such interactions, we determine the average time during which  the antibodies remain close to each other. Their life time is calculated by means of an auto-correlation function as
%\fc{check later that symbols are consistent}
\begin{equation}
     \upsilon=\frac{N_\upsilon(t)N_\upsilon(0)}{N_\upsilon(0)^2}
 \end{equation}
where $N_\upsilon(t)$ is the number of interactions occurring in the system within $r_0$ at time $t$ and $N_\upsilon(0)$ is the same quantity for an initial configuration considered.
Since it is likely that antibodies fluctuate in the vicinity of their neighbours, we considered the interaction to be lost only in case it is not re-established within a time window $\Delta t=36$ and $72\tau$ for implicit and explicit ions models, respectively. We have verified that the obtained results do not qualitatively dependent on the particular choice of this time window.

%\showmatmethods{}% Display the Materials and Methods section

\section*{Acknowledgments}
We thank Susana Marìn-Aguilar, Lorenzo Rovigatti and Letizia Tavagnacco for helpful discussions. F.C. acknowledges funding from the European Union’s Horizon Europe research and innovation program under the grant agreement number 101106720, Marie Skłodowska-Curie Action (MSCA) Postdoctoral Fellowship, project many(Anti)Bodies - mAB. We gratefully acknowledge the financial support by the Swedish Research Council (VR; Grant Nos. 2019-06075 and 2022-03142). 
The computer simulations were also enabled by resources provided by the National Academic Infrastructure for Supercomputing in Sweden (NAISS) and the Swedish National Infrastructure for Computing (SNIC) at Lund University, partially funded by the Swedish Research Council through grant agreements no. 2022-06725 and no. 2018-05973. 
This work is also part of the “LINXS Antibodies in Solution” research program and we acknowledge the financial support by the LINXS Institute of Advanced Neutron and X-ray Science.

\section*{Author contributions}

Author contributions are defined based on CRediT (Contributor Roles Taxonomy). Conceptualization: F.C., E.Z., P.S.; Formal analysis: F.C.; Funding acquisition: F.C., A.S., P.S.; Investigation: F.C., M.P., A.S., E.Z., P.S.; Methodology: F.C., E.Z.; Project administration: F.C., P.S.; Software: F.C.; Supervision: P.S.; Validation: F.C.; Visualization: F.C.; Writing – original draft: F.C., E.Z., P.S.; Writing – review and editing: F.C., M.P., A.S., E.Z., P.S..

\clearpage
\newpage
%\widetext
%\begin{widetext}
%\begin{center}
\onecolumngrid
\begin{center}
\large
\textbf{Electrostatics and viscosity are strongly linked\\in concentrated antibody solutions\\ \bigskip Supplementary Material}

\normalsize
\bigskip
Fabrizio Camerin\textsuperscript{ 1}, Marco Polimeni\textsuperscript{ 1}, Anna Stradner\textsuperscript{ 1, 2},\\ Emanuela Zaccarelli\textsuperscript{ 3, 4}, Peter Schurtenberger\textsuperscript{ 1, 2}\\
\medskip
\small
\textit{%
\textsuperscript{1}Division of Physical Chemistry, Department of Chemistry, Lund University, Lund, Sweden\\
\textsuperscript{2}LINXS Institute of Advanced Neutron and X-ray Science, Lund University, Lund, Sweden\\
\textsuperscript{3}CNR Institute of Complex Systems, Uos Sapienza, Piazzale Aldo Moro 2, 00185 Roma, Italy\\
\textsuperscript{4}Department of Physics, Sapienza University of Rome, Piazzale Aldo Moro 2, 00185 Roma, Italy
}

\end{center}

\normalsize
\bigskip
%\twocolumngrid
%\renewcommand{\thefigure}{S\arabic{figure}}\setcounter{figure}{0}
\renewcommand{\theequation}{S\arabic{equation}}\setcounter{equation}{0}
\renewcommand{\thefigure}{S\arabic{figure}}\setcounter{figure}{0}
\renewcommand{\thetable}{S\arabic{table}}\setcounter{table}{0}

\section*{Antibody atomistic structure}

\begin{figure}[h!]
\centering
\includegraphics[width=0.25\textwidth]{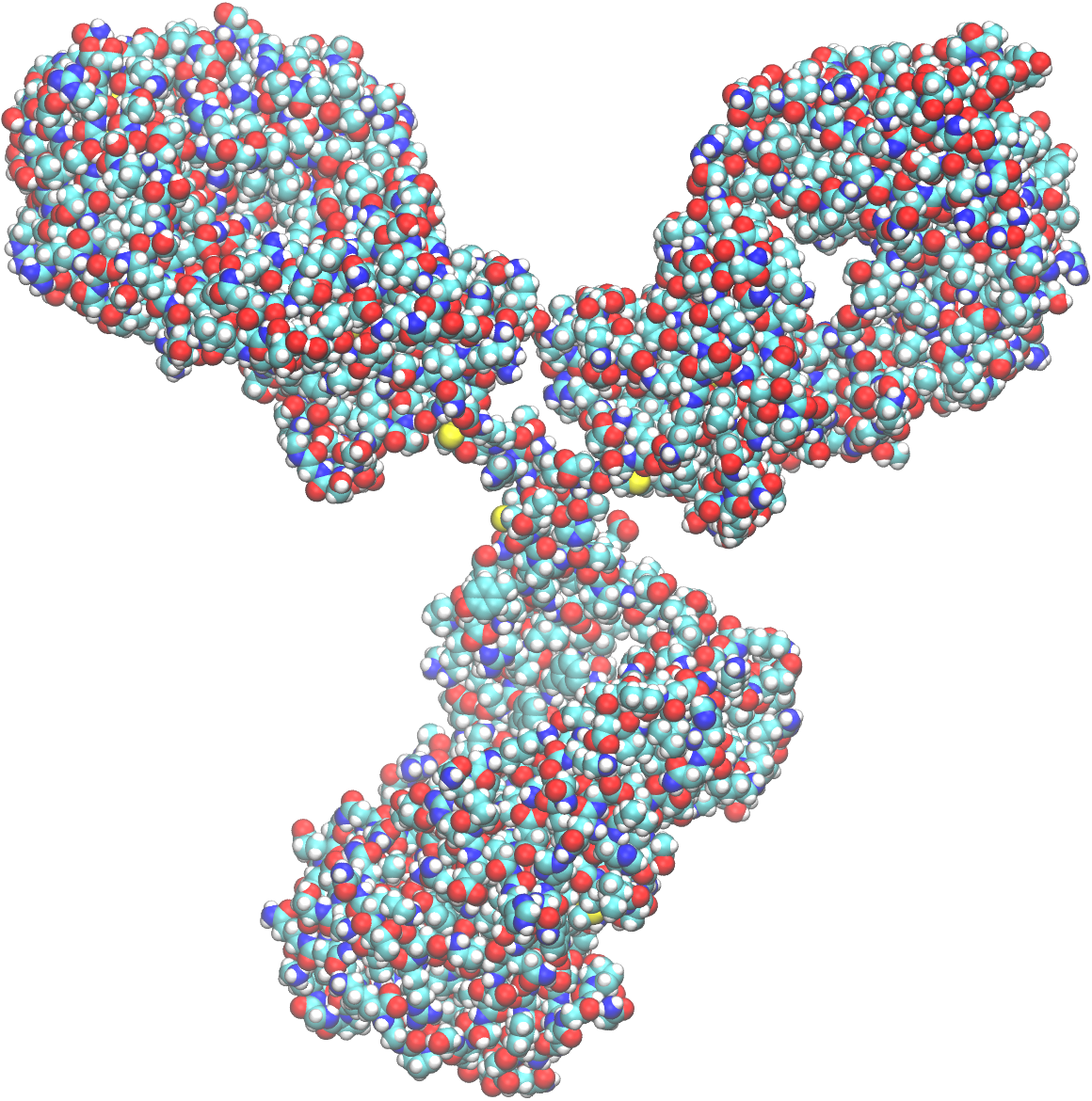}
\caption{\small \textbf{All-atom antibody.} Atomistic representation of the monoclonal antibody under investigation.}
\label{fig:atomistic}
\end{figure}

\clearpage
\newpage

\section*{Antibody isopotential surfaces}

\begin{figure}[h!]
\centering
\includegraphics[width=0.9\textwidth]{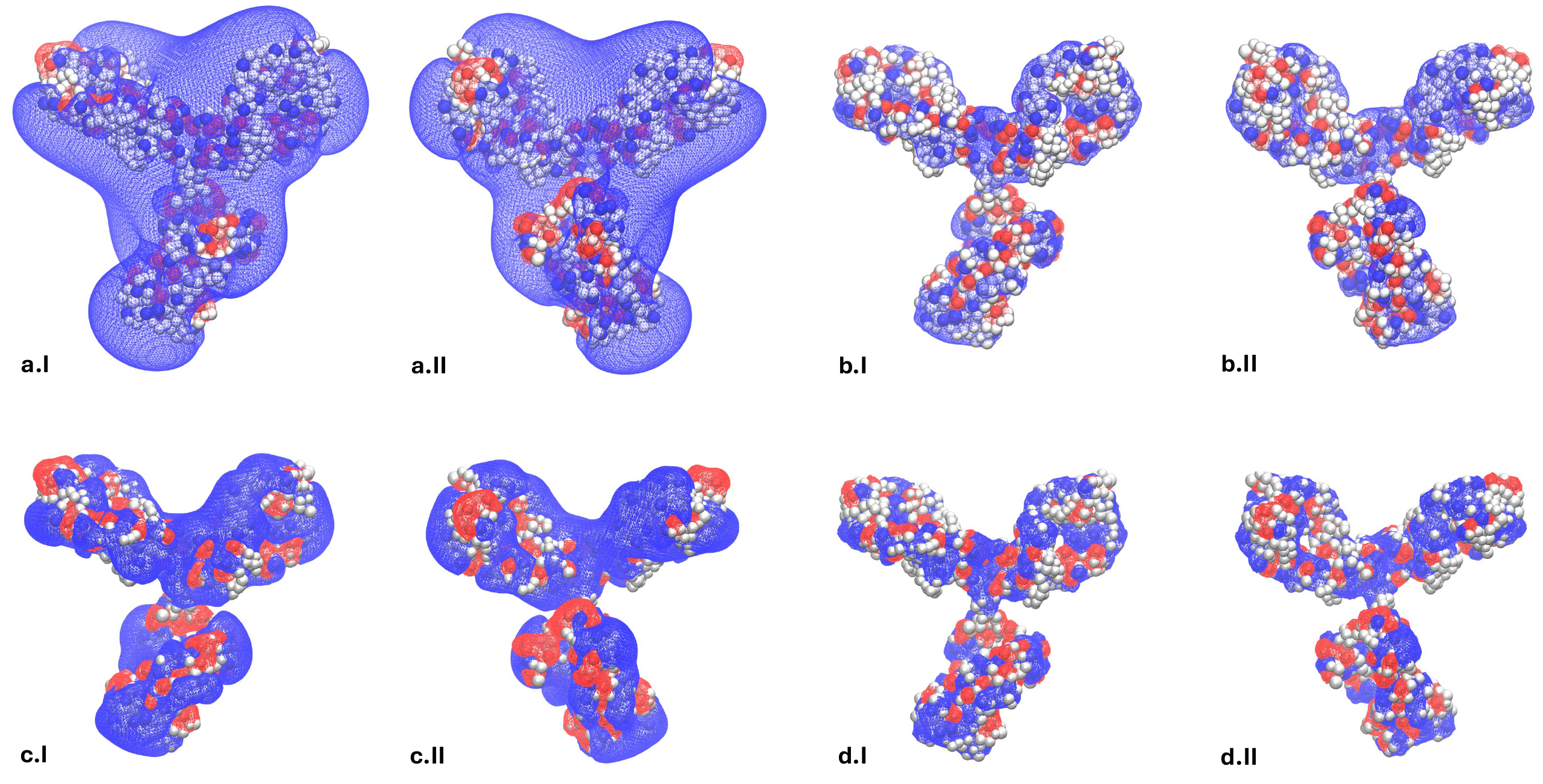}
\caption{\small \textbf{Antibody electrostatic iso-surfaces.} Antibody electrostatic iso-potential surfaces for ionic strengths (a,b) $I=7$ and (c,d) $57$ mM related to iso-values of (a,c) $1 k_{B}T/e$ and (b,d) $3 k_{B}T/e$. Blue wireframe surfaces indicate positive values of the potential, whereas red is for negative values. The iso-surfaces are calculated using the Adaptive Poisson-Boltzmann Solver (APBS)~\cite{jurrus2018improvements} and visualized using VMD~\cite{humphrey1996vmd}. In transparency, we also report the antibody coarse-grained representation where each bead can be positively charged (blue), negatively charged (red), or neutral (gray) to reflect the condition at the respective ionic strength, as calculated from constant-pH Monte Carlo simulations (see Methods). Panels I and II shows the same antibody condition with a view rotated by $180^{\circ}$, respectively.}
\label{fig:isosurfaces}
\end{figure}

\clearpage
\newpage

\section*{Static structure factors for the implicit ions models}

\begin{figure}[h!]
\centering
\includegraphics[width=1\textwidth]{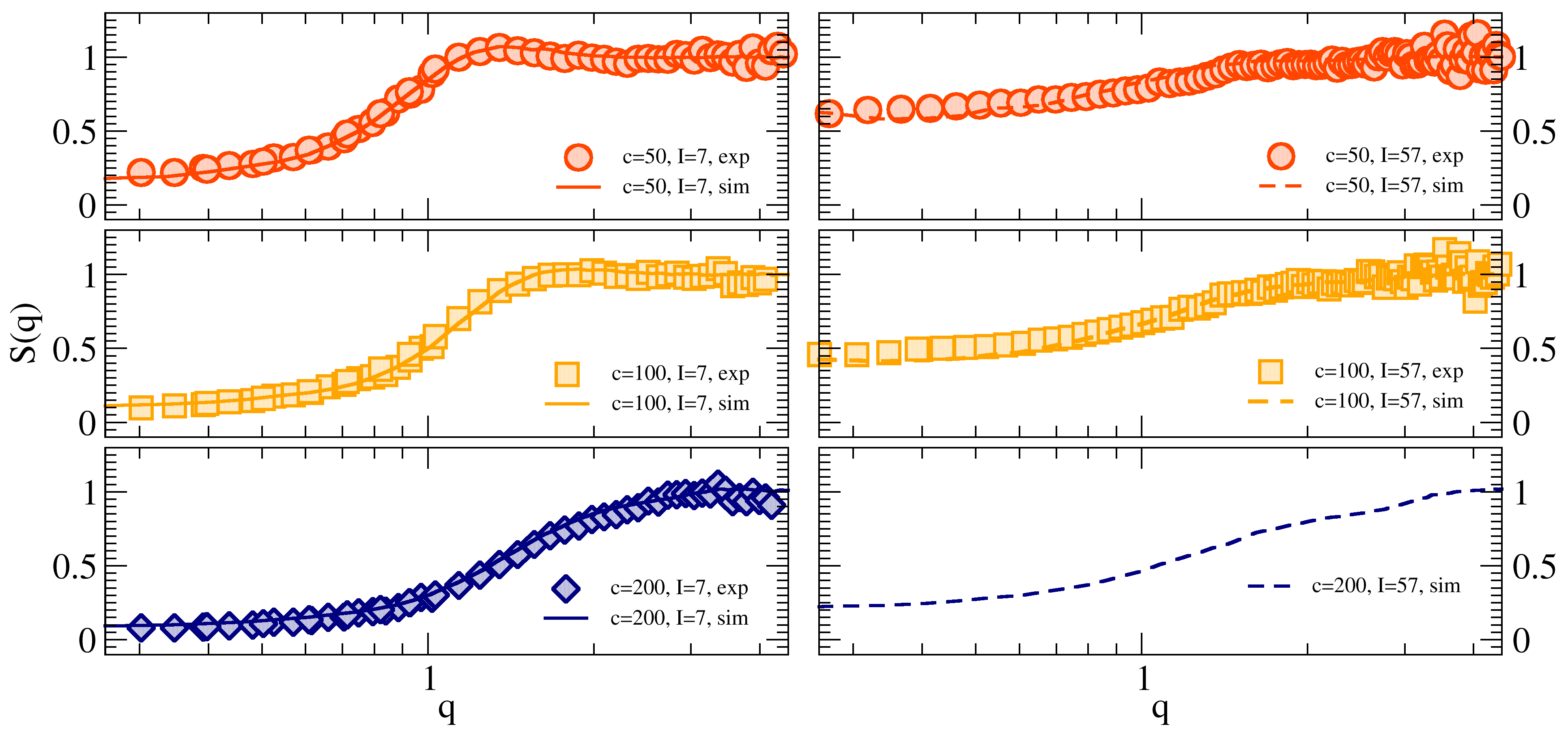}
\caption{\small \textbf{Static structure factors for the implicit ions model with  uniform charge distribution.} Static structure factor $S(q)$ as a function of the wavenumber $q$ for ionic strengths (a) $I=7$ mM and (b) $I=57$ mM for different antibody concentrations $c$, as from the legend, for simulations (full and dashed lines, respectively) and experiments (symbols).}
\label{fig:msd_exp}
\end{figure}

\begin{figure}[h!]
\centering
\includegraphics[width=1\textwidth]{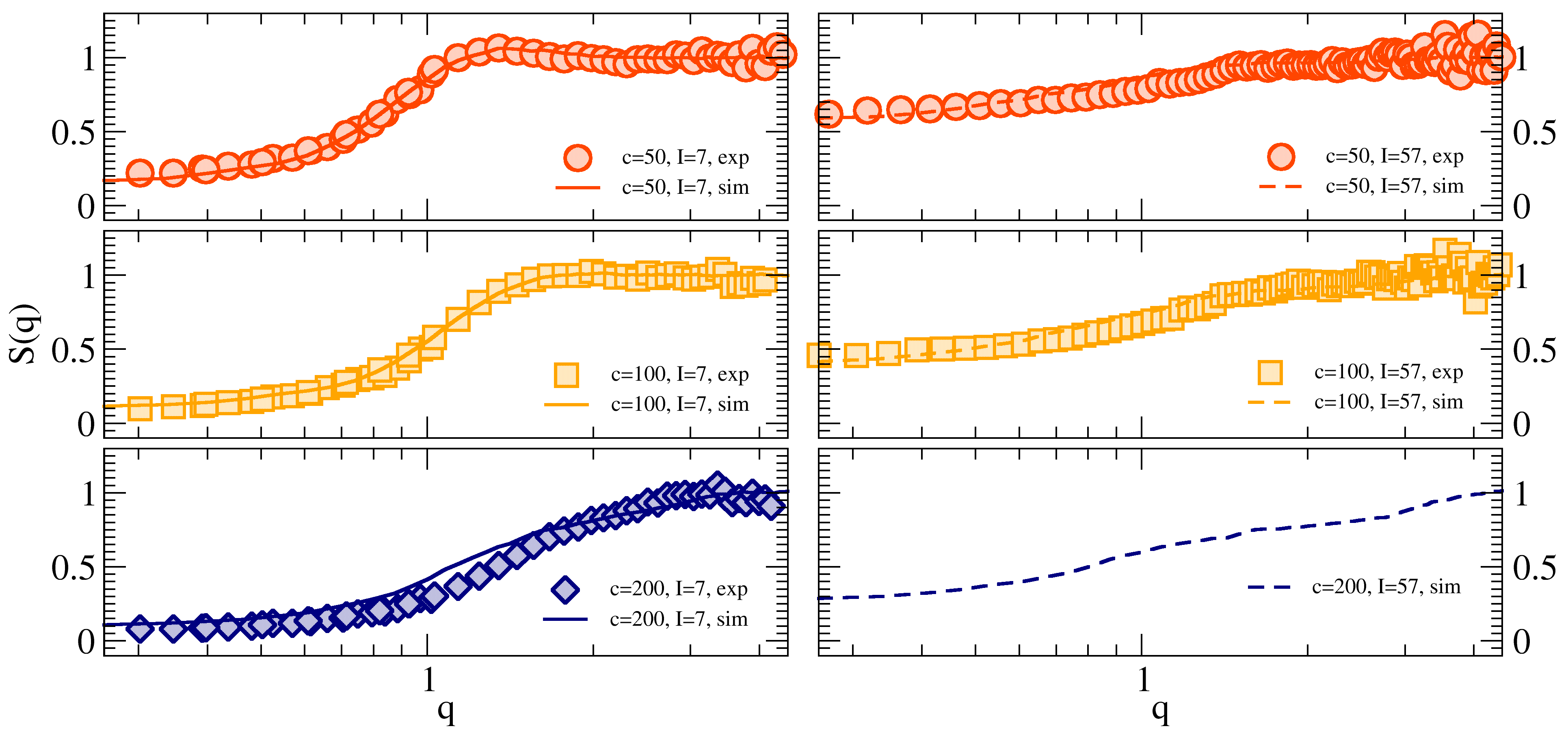}
\caption{\small \textbf{Static structure factors for the implicit ions model with  heterogeneous charge distribution.} Static structure factor $S(q)$ as a function of the wavenumber $q$ for ionic strengths (a) $I=7$ mM and (b) $I=57$ mM for different antibody concentrations $c$, as from the legend, for simulations (full and dashed lines, respectively) and experiments (symbols).}
\label{fig:msd_exp}
\end{figure}

\clearpage
\newpage

\section*{Stress autocorrelation function for other concentrations for implicit ions model}

\begin{figure}[h!]
\centering
\includegraphics[width=\textwidth]{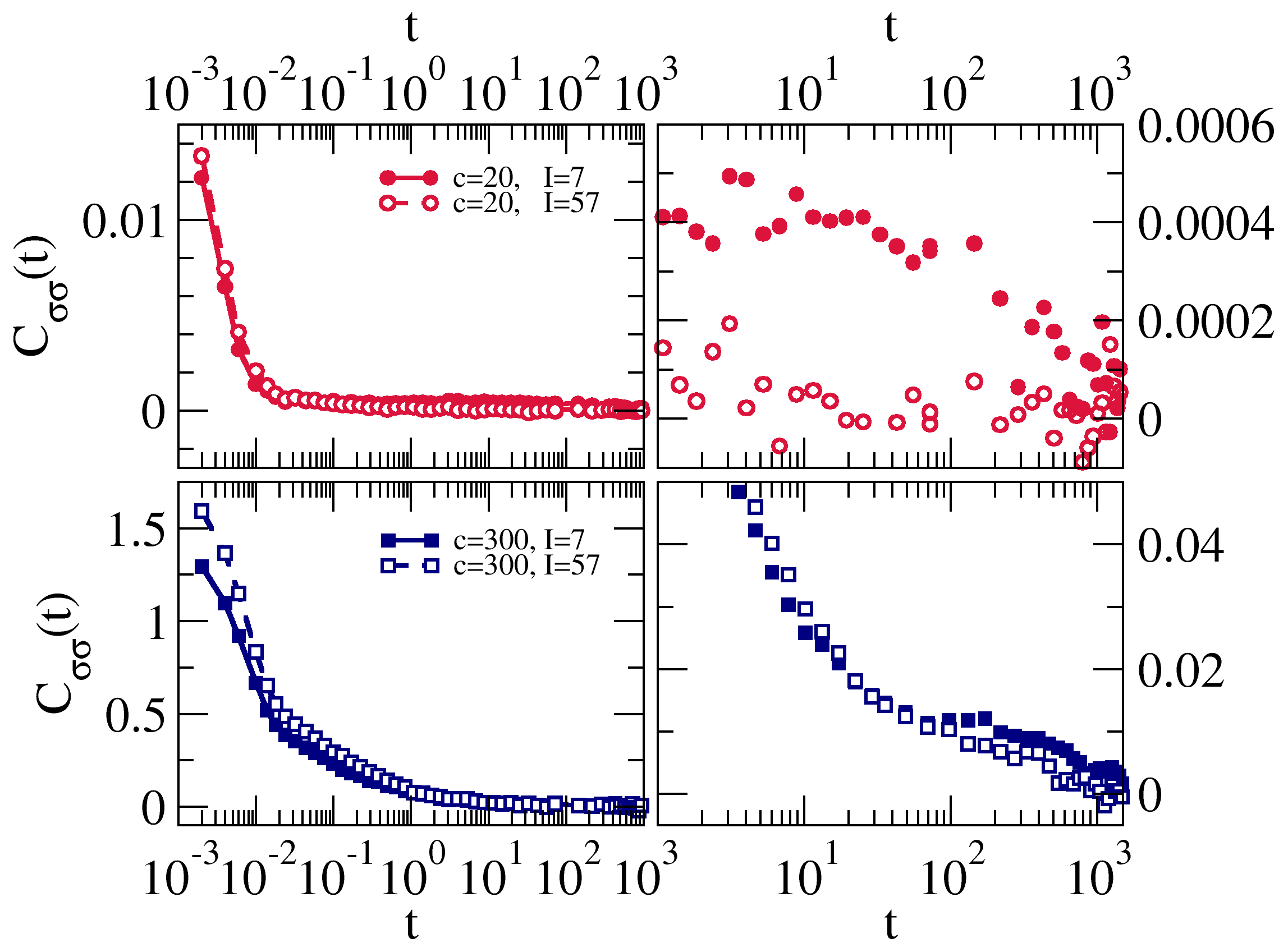}
\caption{\small \textbf{Rheological response for the uniform implicit ions model.} Stress autocorrelation function $C_{\sigma\sigma}(t)$ as a function of simulation time $t$ for the implicit ions model with uniform charge distribution for ionic strengths $I=7$ mM (filled symbols) and $I=57$ mM (empty symbols) for different antibody concentrations $c$ (expressed in mg/ml), as from the legends. Panels to the right show an enlargement of $C_{\sigma\sigma}(t)$ for the long-time correlations.}
\label{fig:msd_exp}
\end{figure}

\begin{figure}[h!]
\centering
\includegraphics[width=\textwidth]{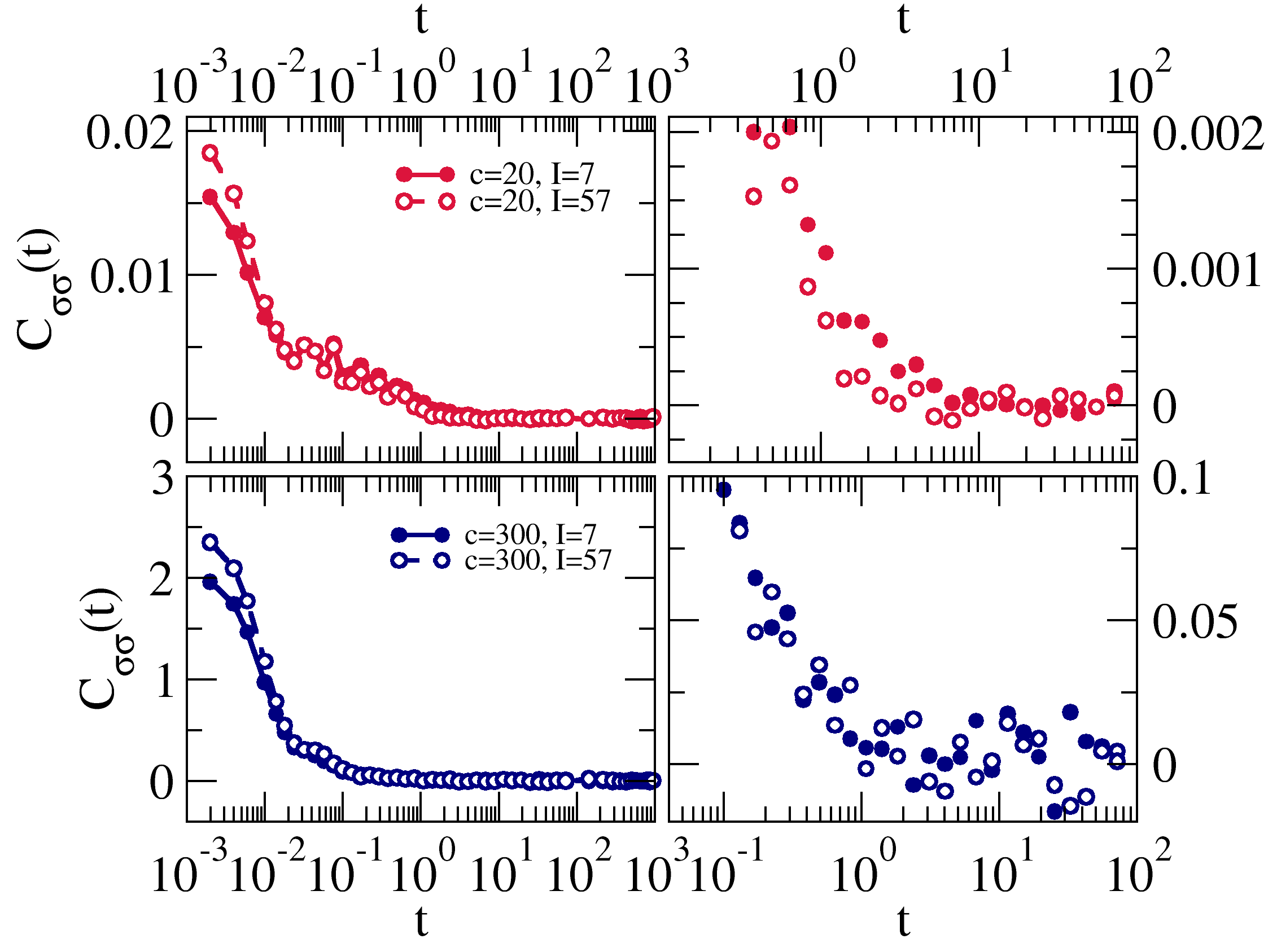}
\caption{\small \textbf{Rheological response for the heterogeneous implicit ions model.} Stress autocorrelation function $C_{\sigma\sigma}(t)$ as a function of simulation time $t$ for the implicit ions model with heterogeneous charge distribution for ionic strengths $I=7$ mM (filled symbols) and $I=57$ mM (empty symbols) for different antibody concentrations $c$ (expressed in mg/ml), as from the legends. Panels to the right show an enlargement of $C_{\sigma\sigma}(t)$ for the long-time correlations.}
\label{fig:msd_exp}
\end{figure}

\clearpage
\newpage

\section*{Viscosity $\eta$ for the explicit ions model}

\begin{figure}[h!]
\centering
\includegraphics[width=0.4\textwidth]{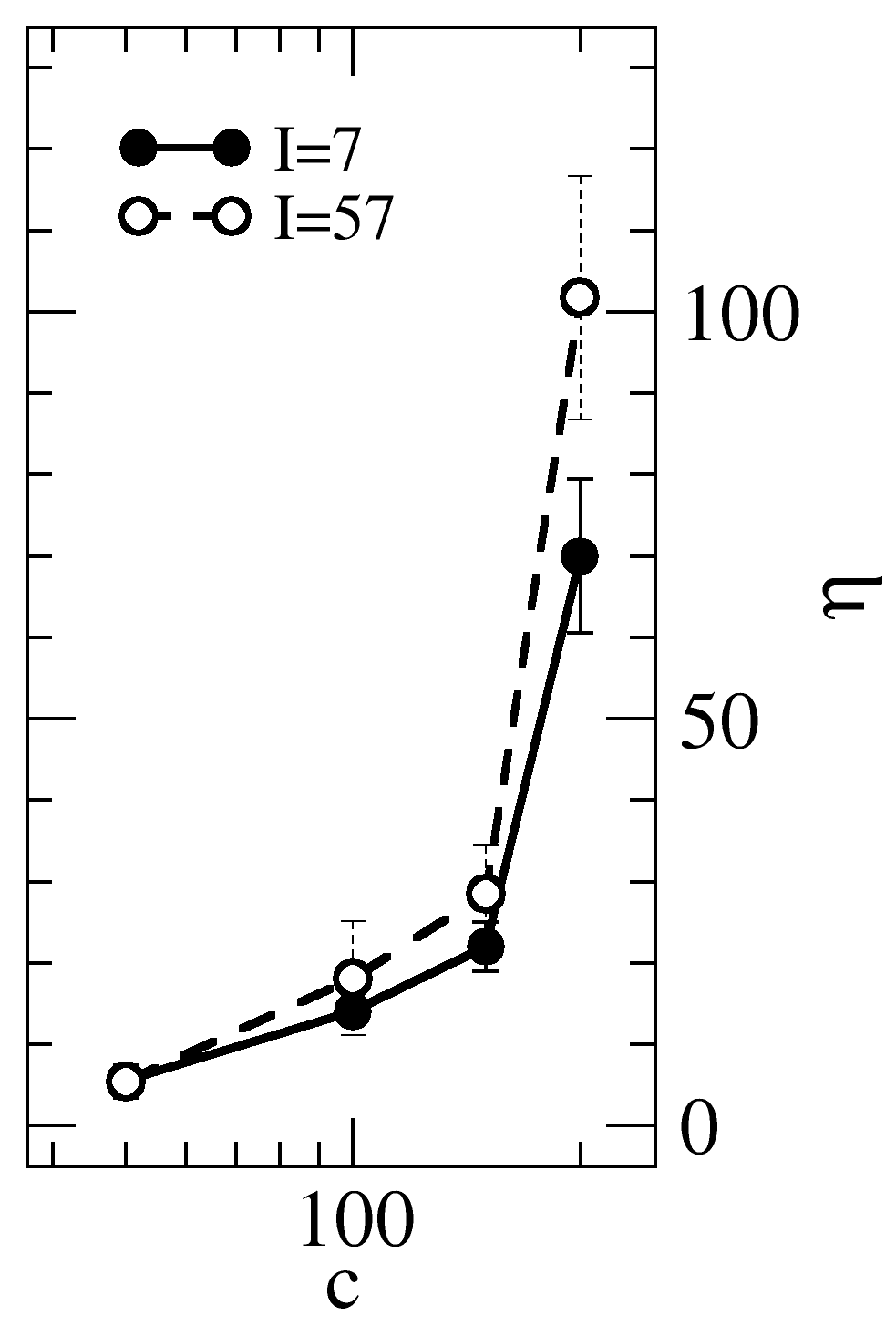}
\caption{\small \textbf{Viscosity for the explicit ions model.} Viscosity $\eta$ as a function of the antibody solution concentration $c$ extracted from simulations with a model with a heterogeneous charge distribution and treated explicitly for two ionic strengths $I=7$ and $57$ mM.}
\label{fig:msd_exp}
\end{figure}

\clearpage
\newpage

\section*{Stress autocorrelation function for models with no attraction or attraction only}

\begin{figure}[h!]
\centering
\includegraphics[width=\textwidth]{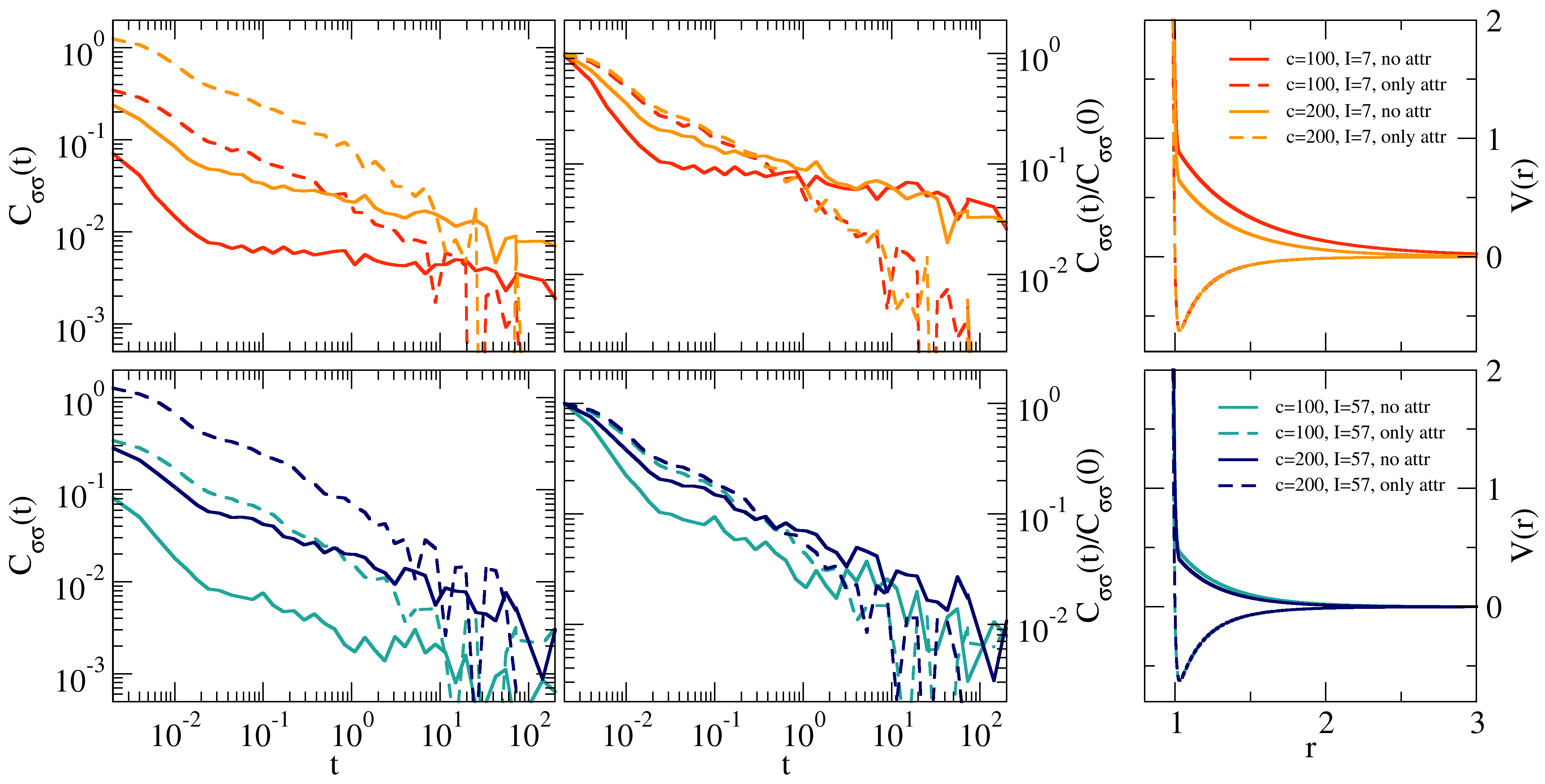}
\caption{\small \textbf{Rheological response for specifically designed interaction potentials.} (a,b) Stress autocorrelation function $C_{\sigma\sigma}(t)$ and normalized stress autocorrelation function $C_{\sigma\sigma}(t)/C_{\sigma\sigma}(0)$ as a function of simulation time $t$ for ionic strengths $I=7$ mM (full lines) and $I=57$ mM (dashed lines) for different antibody concentrations $c$, as from the legends. Panels on the right report the interaction potentials used for each of the cases analysed. Data are for simulations with the effective implicit charged model.}
\label{fig:msd_exp}
\end{figure}

\clearpage

\section*{Static structure factors for models with different angles among the arms}

\begin{figure}[h!]
\centering
\includegraphics[width=0.6\textwidth]{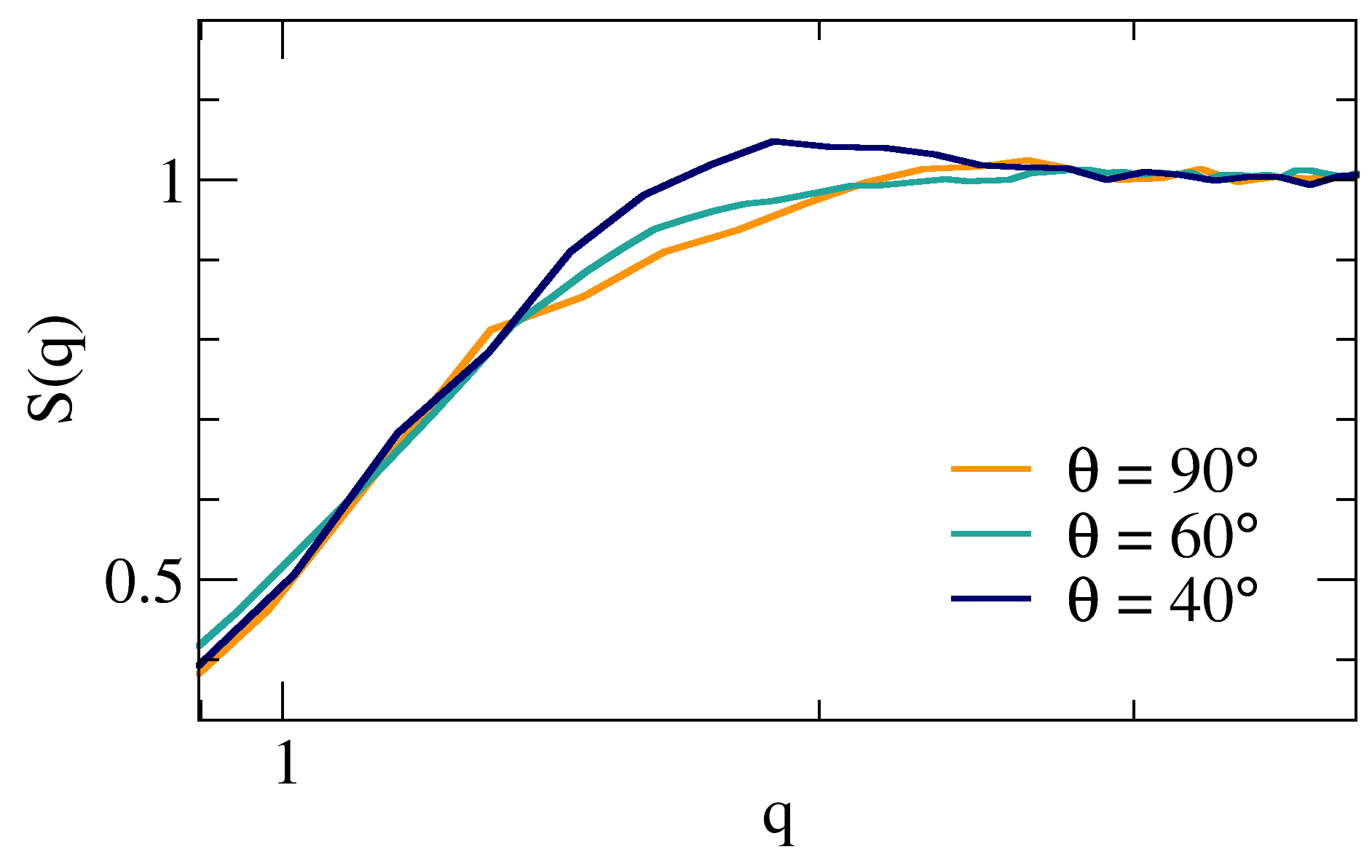}
\caption{\small \textbf{Static structure factors obtained for antibody models with different angles.} Static structure factor $S(q)$ as a function of the wavenumber $q$ for ionic strength $I=7$ mM and antibody concentration $c=100$ mg/ml for simulations run with different antibody models in which the angle between the upper arms $\theta$ (see Figure 1 of the main text) is $90^\circ, 60^\circ$ and $40^\circ$. The outcomes presented in the main text are for $\theta=60^\circ$. Data are for simulations with an effective implicit interaction potential.}
\label{fig:msd_exp}
\end{figure}

\clearpage

\section*{Microscopic characterization}

\begin{figure}[h!]
\centering
\includegraphics[width=\textwidth]{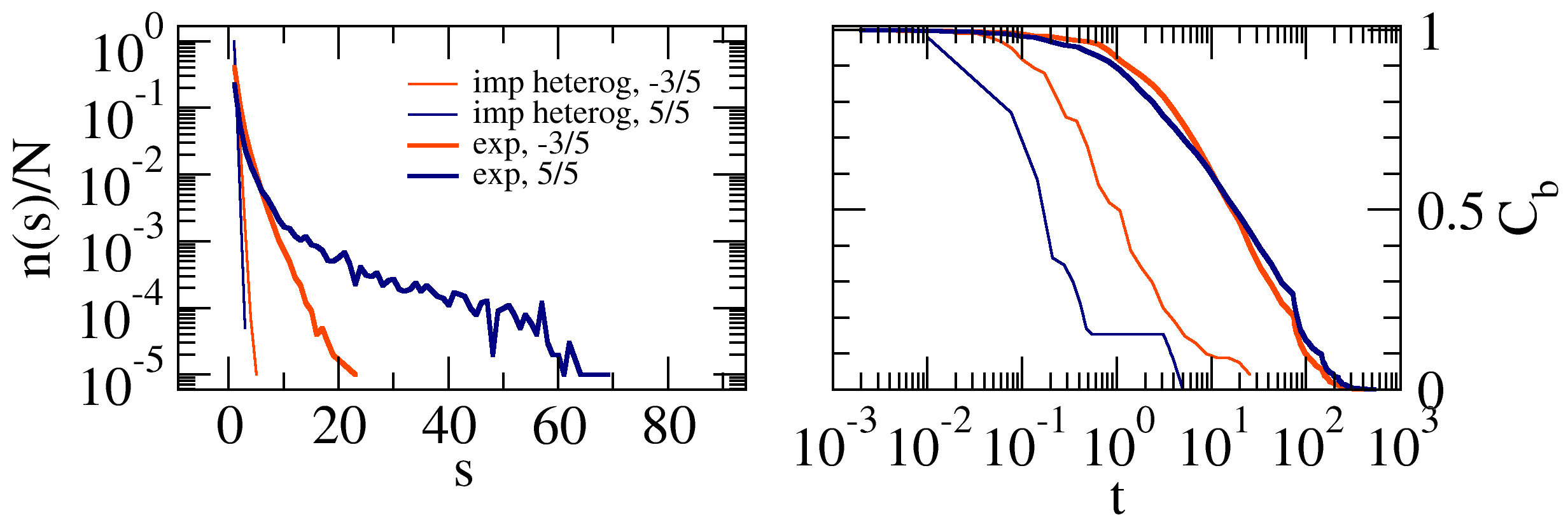}
\caption{\small \textbf{Transient interacting antibodies.} (Left) Size distribution $n(s)$ of transient aggregates of antibodies of size $s$, (right) bond correlation function $C_b$ as a function of simulations time $t$ for the implicit ions model with heterogeneous charge distribution (thin lines) and for the explicit ions model (thick lines), for the interaction between beads with charges $-3/5$ (orange lines) and $5/5$ (blue lines), for a concentration $c=200$ mg/ml and ionic strength $I=7$ mM.}
\label{fig:msd_exp}
\end{figure}

\begin{figure}[h!]
\centering
\includegraphics[width=\textwidth]{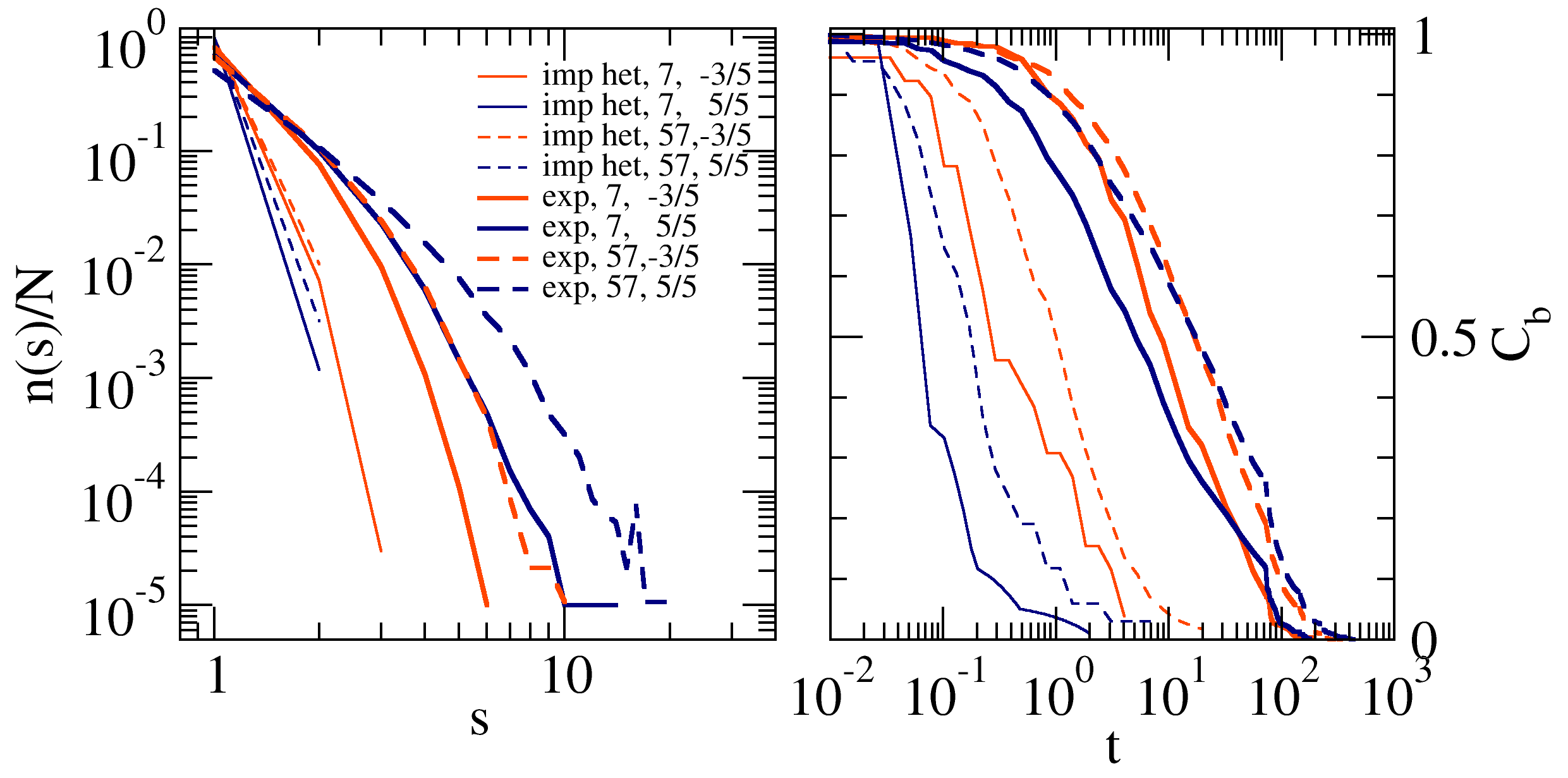}
\caption{\small \textbf{Transient interacting antibodies.} (Left) Size distribution $n(s)$ of transient aggregates of antibodies of size $s$, (right) bond correlation function $C_b$ as a function of simulations time $t$ for the implicit ions model with heterogeneous charge distribution (thin lines) and for the explicit ions model (thick lines), for the interaction between beads with charges $-3/5$ (orange lines) and $5/5$ (blue lines), for a concentration $c=100$ mg/ml and ionic strength $I=7$ (full lines) and $57$ (dashed lines) mM, as from the legend.}
\label{fig:msd_exp}
\end{figure}

\begin{figure}[t!]
\centering
\includegraphics[width=0.5\textwidth]{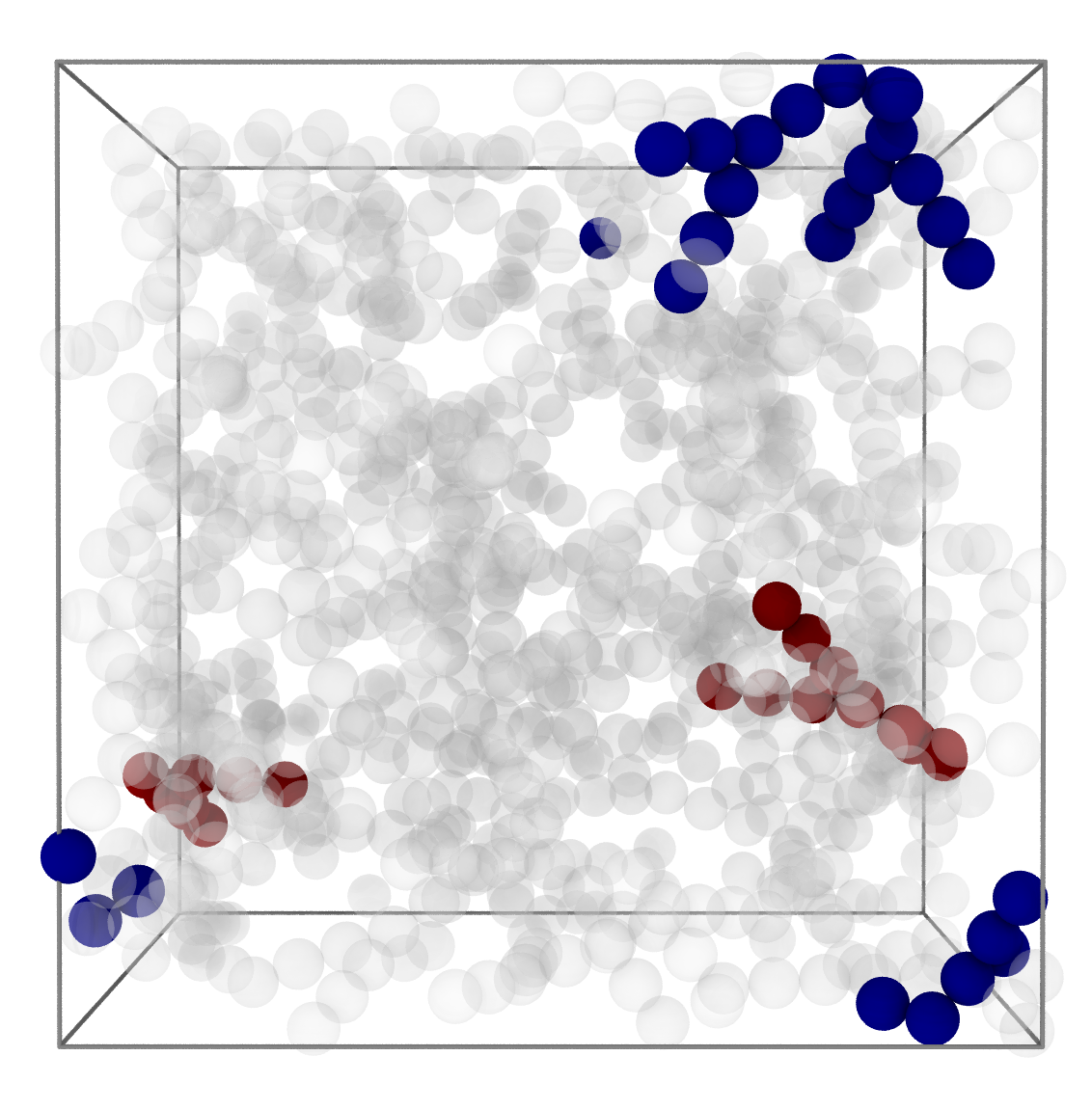}
\caption{\small \textbf{Transient interacting antibodies.} Representative simulation snapshot showing two transient aggregates with $s=2, 3$ for the implicit ions model with heterogeneous charge distribution for $c=200$ and $I=57$ mM. Individual antibodies are shown in transparency.}
\label{fig:msd_exp}
\end{figure}

\twocolumngrid
\clearpage
\newpage

\bibliography{./main_arxiv}

\end{document}